\documentclass[prb,showpacs,floatfix,epsf,epsfig]{revtex4} 
\usepackage{graphicx}
\begin{document}
\title{Half-metallicity versus Symmetry in Pt, Ni and Co-based Half Heusler Alloys: A First-principles Calculation}
\author{Madhusmita Baral$^{1,2}$ and Aparna Chakrabarti$^{2,3}$}
\affiliation{$^{1}$ Synchrotrons Utilization Section, Raja Ramanna Centre for Advanced Technology, Indore - 452013, India}
\affiliation{$^{2}$ Homi Bhabha National Institute, Training School Complex, Anushakti Nagar, Mumbai-400094, India}
\affiliation{$^{3}$ Theory and Simulations Laboratory, HRDS, Raja Ramanna Centre for Advanced Technology, Indore - 452013, India}

\begin{abstract}

Using first principles calculations based on density functional
theory, we study the geometric, electronic, and magnetic
properties of Pt, Ni and Co-based half Heusler alloys, namely, 
 Pt$BC$, Ni$BC$ and Co$BC$ ($B$ = Cr, Mn and Fe; $C$ = Al, Si, P, S,  
Ga, Ge, As, Se, In, Sn, Sb and Te). We calculate the formation energy
of these alloys in various crystal symmetries, which include, 
the (face-centered) cubic $C1_{b}$ ($F\bar{4}$3m), 
orthorhombic ($Pnma$), 
as well as hexagonal ($P\bar{6}2m$ and $P6_{3}/mmc$) structures. It 
 has been observed that out of all the 108 structures, studied here,  
energetically stable cubic structure is observed for only 18 
materials. These alloys are primarily 
having either a $C$ atom or an $A$ atom with a high atomic number. 
We also observe that along with the alloys with $C$ atoms from group 
IIIA, IVA and VA -- alloys with $C$ atoms from group VIA are also 
found to be, by and large, energetically stable. 
To examine the relative stabilities of different symmetries  
in order to search for the respective lowest energy state for 
each of the above-mentioned systems, as well as to find whether a 
material in the ground state is half-metallic or not, we analyze the 
formation energy, and the electronic density of states, in detail. 
Based on these analyses, the possibility of existence of 
any {\it one-to-one relationship} 
between the {\it cubic symmetry} and the 
{\it half-metallicity} in these half Heusler alloys is probed. 
Subsequently, we predict about the existence of a few new 
{\it non-cubic} half Heusler alloys with 
substantially low density of states at one of the spin channels 
and reasonably {\it high spin polarization at the Fermi level}. 
\end{abstract}


\maketitle

\section{Introduction}

The prediction and development of new half-metallic ferromagnets are
of immense interest due to their potential for technological 
application.\cite{generalintro} The half-metallic (HM) ferromagnets 
(FM) are a type of FM material where spin polarization at the Fermi 
level (E$_{F}$) is high (expected to be 100\%) which may have an 
application in the field of spintronics. Ever since the 
half-metallicity has been 
predicted, in half Heusler alloys (HHA), namely, NiMnSb and related
 isoelectronic compounds, PtMnSb and PdMnSb, 
on the basis of band structure calculations,\cite{NiMnSbasHM}  
the field of half-metallic half Heusler alloys (HM HHA) has 
attracted the interest of the researchers.\cite{PRB95JM}
Here we wish to point out that, further on, we refer to all the 
materials with high (about 65 to 70\%) to 100\% spin polarization,
as HM-like materials.

Many HM-like materials have been found theoretically, among the
 half and full Heusler alloys.\cite{NiMnSbasHM,PRB93TR,JMMM423TR} 
 The Curie temperature (T$_{C}$) 
of quite a few of these alloys is calculated to be higher than the 
room temperature which is essential for their application as an
efficient and useful spin-injector material.\cite{highTCHM}
For this application, the interest is indeed in magnetic 
HHAs. It has been seen that a large amount of work on NiMnSb and 
substitution at its different atomic sites 
 has been carried out, both theoretically and
experimentally.\cite{NiMnSbasHM,workonNiMnSb} In order to search for 
new half-metals, a large number of general $ABC$ type HHAs have also 
been studied in the literature, where $A$ and $B$
 are transition metal atoms and $C$ is an sp element.\cite{workonABC} 

In the literature, most of the HHAs studied theoretically have been 
 shown to possess half-metallic property in the cubic C1$_{b}$ 
phase with F$\bar{4}$3m space-group.\cite{NiMnSbasHM} However, 
only some of these have been experimentally 
synthesized.\cite{exptHMHHAcubic} On the contrary, it has been seen 
in the literature that many of the half Heusler alloy samples 
exhibit non-cubic symmetries but there is no explicit discussion
on the half-metallicity in these materials.\cite{workon-non-cubic-Co+Ni-Cr+Mn+Fe-Si,workon-non-cubic-Co-Cr+Mn+Fe-P+As++Ni-Cr+Mn+Fe-P+As,workon-non-cubic-Co+Ni-Cr+Mn+Fe-P++Co+Ni-Cr+Mn-As,workon-non-cubic-CoCrGe,workon-non-cubic-Co+Ni-MnGe,workon-non-cubic-CoMnGe++Co+Ni-FeGe,workon-non-cubic-Co+Ni-Mn+Fe-P,workon-non-cubic-CoMnAs,workon-non-cubic-CoMnSb,workon-non-cubic-NiCrP,workon-non-cubic-NiMnGa,workon-non-cubic-NiMnGe,workon-non-cubic-Pt-Cr+Fe-Sn,workon-non-cubic-PtMnGe++PtCrSb,JMMM38}
It has been observed that 
NiMnSb and some other $AB$Sb, as well as $AB$Sn materials 
  possess cubic C1$_{b}$ symmetry.\cite{JMMM38} 
On the other hand, while NiMnGe is seen to exhibit an orthorhombic 
$Pnma$ space-group at room temperature,\cite{JMMM27} NiMnGa has a 
hexagonal symmetric structure.\cite{workon-non-cubic-NiMnGa}
Further, $AB$Si or $AB$Ge or $AB$P or $AB$As compounds are seen 
to exhibit either an 
orthorhombic structure (with $Pnma$ space-group) 
or a hexagonal phase (with either $P6_{3}/mmc$ or
$P\bar{6}2m$ space-group).\cite{workon-non-cubic-Co+Ni-Cr+Mn+Fe-Si,workon-non-cubic-Co-Cr+Mn+Fe-P+As++Ni-Cr+Mn+Fe-P+As,workon-non-cubic-Co+Ni-Cr+Mn+Fe-P++Co+Ni-Cr+Mn-As,workon-non-cubic-CoCrGe,workon-non-cubic-Co+Ni-MnGe,workon-non-cubic-CoMnGe++Co+Ni-FeGe,workon-non-cubic-Co+Ni-Mn+Fe-P,workon-non-cubic-CoMnAs,workon-non-cubic-CoMnSb,workon-non-cubic-NiCrP,workon-non-cubic-NiMnGa,workon-non-cubic-NiMnGe,workon-non-cubic-Pt-Cr+Fe-Sn,workon-non-cubic-PtMnGe++PtCrSb}
In the literature, there are also reports that, many samples of HHAs 
are seen to exhibit more than one symmetries. 
 For example, each of NiMnAs and NiMnP are reported to 
possess both orthorhombic and hexagonal 
structures.\cite{workon-non-cubic-Co-Cr+Mn+Fe-P+As++Ni-Cr+Mn+Fe-P+As} 
Further, in the literature, it has been found that 
 some materials, including CoMnSb\cite{JMMM70,PRB74VK} and 
PtMnAl\cite{JLM99}, exhibit 
site disorder, which is often associated with a 
larger effective unit cell. As for the magnetic
properties, many of the HHAs studied 
in the literature are seen to be FM in nature, and 
it is observed that with the change in $A$ and $B$ elements, the 
magnetic property of the alloys changes as well.\cite{workonFMHHA}

In the literature, a large number of full Heusler alloys (FHA) as well
 as HHAs, has been studied which shows half-metallic-like behavior 
 but, to the best of our knowledge, only those alloys are seen to 
have a {\it half-metallic-like} character which possess a {\it cubic}
structure. Therefore, in this work, we aim to search and screen 
magnetic half-metals based on half Heusler alloys: our 
specific interest is to probe that whether the cubic symmetry is a 
{\it necessary but NOT sufficient condition for the half-metallicity} 
in the HHA materials. It is to be further noted that,
 full Heusler alloys are much more extensively studied, 
compared to the HHAs, but none of these FHAs is seen to exhibit a 
structure with hexagonal symmetry. On the contrary, it is 
seen that many HHAs exist in different crystal symmetries, which 
include, hexagonal structures (with space-groups $P\bar{6}2m$ and 
$P6_{3}/mmc$) as well as (face-centered) cubic $C1_{b}$ ($F\bar{4}3m$ 
space-group) and orthorhombic structures (with space-group $Pnma$). 
For example, while NiMnSb is a well-known cubic 
HHA, the change in $C$ atom from Sb to As leads to an orthorhombic 
structure in the lowest energy state. There are other similar
examples among the HHAs.
Since many Ni-based HHAs are known to exist in phases other 
than the well-known cubic $C1_{b}$ phase, we aim to study the 
reason as to why there is this difference in stability  
in case of various symmetries in HHAs and if there is
any systematics involved. Further, 
in forming $A_{2}BC$ or $ABC$ Heusler alloys, $C$ atom from the 
p-block elements, of group IIIA, IVA and VA, are well-known in the 
literature. Here we probe whether $C$ atoms from group VIA 
are also favored in forming stable Heusler alloys.

To this end, we choose three sets of compounds, namely, Co$BC$, 
Ni$BC$ and Pt$BC$. Choice of $A$ atom is primarily driven by the 
existing work in the literature, which are based mainly on Ni 
and also some Co and Pt-derived HHA materials.
Further, the $B$ atom is chosen to be an element with high atomic
magnetic moment
keeping the spin-injection properties in mind. It is to be noted that
 we are interested in those alloys, which show both magnetism 
as well as half-metallicity. We study various materials 
in different possible space-groups to 
find and understand the symmetry of the phase with lowest energy
for each of these materials. 
For this purpose, we analyze the formation energy; also partial 
and total density of states (DOS) in order to understand the extent 
of hybridization between different atoms of the alloys. Further, we
 try to understand the trend in similarities and differences in the  
magnetic and electronic properties of these three sets of materials. 
We specifically analyze the possibility of existence of a 
half-metallic-like 
property in these compounds in their lowest energy phases and probe 
if there is any {\it one-to-one relationship} between the 
{\it cubic symmetry} and the {\it half-metallicity} in the alloys, 
studied in this work. Consequently, we predict 
the possibility of a few new half-Heusler alloys 
exhibiting high {\it spin polarization at the Fermi level} (termed as 
$SP$ henceforth) with or without having the cubic symmetry.
In the next section, we discuss the methods of calculations, which
are based on density functional theory. In the section followed by 
methodology, we present our results and discussion on the same.
Finally, we summarize and conclude in the last section.

\section{Method}  

First, we discuss in detail the space-groups we have considered 
in our present work.
We probe four different crystal symmetries, which have been reported
for various half-Heusler alloys so far, namely, 
 cubic $C1_{b}$ (space-group $F\bar{4}3m$, no. 216), 
orthorhombic (space-group $Pnma$, no. 62), as well as 
hexagonal structures (space-group $P\bar{6}2m$, no. 189 and 
  $P6_{3}/mmc$, no. 194). We have not carried out any calculations
on any disordered structure in this paper due to the lack of any 
systematic input of structural data for the materials, studied here.
 In this paper, we have carried out calculations on Co, 
Ni and Pt-based systems; Co, Ni and Pt are taken as $A$ atom. 
Cr, Mn and Fe have been considered as the $B$ atom since we 
are interested in magnetic alloys and these $B$ atoms are known to
have high atomic moments. Further, 
for $C$ atom, we have taken the following elements, Al, Si, P, S, 
Ga, Ge, As, Se, In, Sn, Sb and Te. In total, we have studied
108 different HHAs. We probe the symmetry of an alloy which has the 
lowest formation energy. Further, we have carried out
in depth calculations of the electronic and magnetic properties of
 the energetically stable alloys. The half-Heusler alloys
assume an ordered $ABC$ structure, where the $A$ and $B$ atoms are 
elements with d-electrons, typically transition metal (TM) atoms    
and $C$ atoms are elements with s,p electrons (termed as sp element).

In the lowest energy state, the most well-studied HHA
{\bf NiMnSb} has a $C1_{b}$ structure that consists 
of four interpenetrating face-centered-cubic (fcc) lattices with 
origin at the fractional positions,
(0.25,\,0.25,\,0.25),
 (0.75,\,0.75,\,0.75)
(0.5,\,0.5,\,0.5), and
(0.0,\,0.0,\,0.0).
We label these sub-lattices as $W$, $X$, $Y$ and $Z$, respectively. 
In $C1_{b}$ structure of NiMnSb, the Ni atoms occupy the $W$  
 sub-lattice and the $X$ sub-lattice remain empty. Further, Mn and 
Sb atoms occupy the $Y$ and $Z$ sub-lattices, respectively. 

{\bf CoMnGe} is a HHA which exhibits an 
  orthorhombic structure with $Pnma$ symmetry. 
Atoms here occupy a Wyckoff position of 4c symmetry.
While each of Co, Mn and Ge atoms has four equivalent atoms with 
fractional coordinates $x$ and $z$ as variables, the $y$ coordinate 
for all the four atoms is 0.25. The symmetry equivalent fractional 
coordinates according to the 4c point-group symmetry are as follows:
$x$,0.25,$z$; -$x$+0.5,0.75,$z$+0.5; -$x$,0.75,-$z$;
$x$+0.5,0.25,-$z$+0.5.

{\bf NiMnGa} assumes a hexagonal structure ($P6_{3}/mmc$ space-group)
 where Ni, Mn and Ga have
two equivalent atoms each. Ni atom occupies sites with
point-group symmetry of 2d (1/3, 2/3, 3/4 and 2/3, 1/3 and 1/4);
Mn atom occupies sites with 2a symmetry (0,0,0 and 0,0,0.5); and Ga
atoms are found at the sites with 2c symmetry (1/3,2/3,1/4 and
2/3,1/3,3/4). 

{\bf NiFeAs} is found in a hexagonal structure ($P\bar{6}2m$ 
space-group). Atom Ni occupies 
the site with 3f point-group symmetry ($x$,0,0; 0,$x$,0;-$x$,-$x$,0);
atom Fe has the preference for a site which has a 3g point-group 
symmetry with fractional coordinates as follows: 
 $x$,0,0.5; 0,$x$,0.5; -$x$,-$x$,0.5.
Atom As occupies two different Wyckoff positions with 1b and 2c 
point-group symmetries with fractional coordinates, (0,0,0) and 
(1/3,2/3,0.5 and 2/3,1/3,0.5), respectively.

The equilibrium lattice constants and fractional coordinates 
of all these alloys have been optimized by doing full 
geometry optimization using Vienna Ab Initio Simulation Package 
(VASP)\cite{VASP} which has been used in combination
with the projector augmented wave (PAW) method.\cite{PAW}  
We have interchanged the Wyckoff positions of the $A$ and $B$ atoms
in case of $P6_{3}/mmc$ and $P\bar{6}2m$ space-groups 
as well as varied the variable fractional coordinates, $x_{A}$
and $x_{B}$ in case of the latter space-group to find the
structure with the lowest formation energy. $x$ and $z$ have 
been varied for all the three atoms $A$, $B$ and $C$ in
case of $Pnma$ symmetry to arrive at the structure which yields 
the lowest formation energy among all. We report in this paper 
the results for the fully optimized geometries of the materials 
for each of the four space-groups mentioned above. 

For exchange-correlation functional, 
generalized gradient approximation (GGA) over the local density 
approximation (LDA) has been used.\cite{PBE} 
We use an optimum energy cutoff of 500 eV for the planewave basis-set.
  The final energies have been calculated with a $k$ mesh 
for which the convergence has been tested. 
 The energy and the force tolerance for our calculations 
were 10 $\mu$eV and 10 meV/\AA, respectively. The mixing or 
formation energies ($E_{form}$) have been calculated\cite{VASP} for 
 probing the energetic stability of a material, using the equation 
$E_{form} = E_{tot} - \Sigma_{i} c_{i}E_{i}$, 
where $i$ denotes different types of atoms present in the unit cell 
of the material and $E_{i}$ is the standard state (bulk) energy of
the corresponding atom, $i$.\cite{VASP} These energies have been
then analyzed to establish the energetic stability of the alloys in 
 different crystal symmetries.
The optimized geometries of the systems are compared with the 
results obtained in the literature, wherever the results are 
available. 
The detailed converged structures (fractional coordinates and
lattice constants) will be reported separately.\cite{MBunpubl} 

For in-depth understanding of the magnetic and electronic  
 properties, we have carried out relativistic
spin-polarized {\it all-electron} calculations for the optimized  
 structures of all the systems.
 These calculations have been performed 
using full potential linearized augmented planewave (FPLAPW)
program\cite{Wien2k} with the generalized gradient approximation 
(GGA) for the exchange correlation functional.\cite{PBE}
For obtaining the electronic properties, the Brillouin zone (BZ) 
integration has been carried out using the tetrahedron method 
with Bl\"ochl corrections.\cite{Wien2k} An energy cut-off for the 
planewave expansion of about 14 Ry is typically used. 
The cut-off for charge density is 
$G_{max}$= 14. The number of $k$ points for the self-consistent 
field cycles in the irreducible BZ is 
 about 300, 600 and 2300 in case of cubic, hexagonal, 
and orthorhombic, respectively.
 The convergence criterion for the total energy $E_{tot}$ is 
about 0.1 mRy per atom. The charge convergence is set to 0.0001. 

\section{Results and Discussion}
\begin{table}
Table 1. Formation energy and Lattice Parameter for Co$BC$, calculated for cubic, hexagonal and orthorhombic structures. $GS_{expt}$ gives the experimentally observed symmetries. \\
\begin{tabular}{|c|c|c|c|c|c|c|c|c|c|} \hline
 Material   &$E_{F\bar{4}3m}$ & $E_{P6_{3}/mmc}$ & $E_{P\bar{6}2m}$ &$E_{Pnma}$ & $a$ &$a,c$ &$a,c$  &$a,b,c $ &$GS_{expt}$    \\
            &(kJ/mol)& (kJ/mol) & (kJ/mol) &(kJ/mol) &  (\AA)     & (\AA)        &(\AA)         &(\AA)        &         \\
            &        &          &          &         & $F\bar{4}3m$     & $P6_{3}/mmc$        &$P\bar{6}2m$         &$Pnma$        &         \\
\hline CoCrAl &   8.85& -31.06  & 25.65 & {\bf -41.50}  & 5.47 & 4.18, 4.77 & 5.98, 3.75  & 4.90, 4.03, 7.34 &--  \\
 CoCrGa &  21.98& -6.48   & 18.46 & {\bf -7.29}   &5.47 & 4.17, 4.84 &6.81, 2.94  &4.94, 4.05, 7.32 & -- \\
 CoCrSi & -57.32& -88.95  &-100.00&{\bf -103.25} & 5.38 & 4.00, 4.99 &5.83, 3.60  & 5.75, 3.61, 6.73   &$Pnma$ \cite{workon-non-cubic-Co+Ni-Cr+Mn+Fe-Si} \\
 CoCrGe & -9.15 & {\bf -23.89}  &-20.48 & -23.76  & 5.49 & 4.10, 5.11 &6.05, 3.68  &5.73,3.84,7.06   & $P6_{3}/mmc$  \cite{workon-non-cubic-CoCrGe} \\
 CoCrP  &-77.89 & -96.42  &-160.98&{\bf -168.72}  & 5.35 & 3.87, 5.17 &5.71, 3.53  &5.73, 3.52, 6.68 &$Pnma$ \cite{workon-non-cubic-Co-Cr+Mn+Fe-P+As++Ni-Cr+Mn+Fe-P+As,workon-non-cubic-Co+Ni-Cr+Mn+Fe-P++Co+Ni-Cr+Mn-As} \\
 CoCrAs &-28.34 & -18.19  & -49.69& {\bf -52.78}  & 5.53 & 4.05, 5.28 &6.08, 3.67  &5.95, 3.71, 6.99  & $P\bar{6}2m$  \cite{workon-non-cubic-Co-Cr+Mn+Fe-P+As++Ni-Cr+Mn+Fe-P+As,workon-non-cubic-Co+Ni-Cr+Mn+Fe-P++Co+Ni-Cr+Mn-As} \\
 CoCrS  & 34.26 & -22.08  &-44.12 & {\bf -47.54}  & 5.46 & 3.63, 6.71 &5.74, 3.57  &5.92, 3.47, 6.73   &-- \\
 CoMnAl &  23.70& -65.94  &-49.45 & {\bf -75.45}  & 5.46 & 4.12, 5.03 &6.86, 2.81  &5.05, 3.99, 7.31  &--\\
 CoMnGa &  41.72& -41.65  &-34.74 & {\bf -43.76}  & 5.47 & 4.12, 5.18 &6.76, 2.98  &5.7, 4.05, 7.24   &--\\
 CoMnSi & -58.02&-118.66  &-123.67&{\bf -129.68}  & 5.38 & 3.97, 5.00 & 5.97, 3.50 &5.72, 3.66, 6.87  &$Pnma$ \cite{workon-non-cubic-Co+Ni-Cr+Mn+Fe-Si} \\
 CoMnGe & -14.32& -53.13  &-55.51 & {\bf -56.65}  & 5.50 & 4.09, 5.13 &6.22, 3.52  &5.83, 3.78, 7.08  &$Pnma$  \cite{workon-non-cubic-Co+Ni-MnGe,workon-non-cubic-CoMnGe++Co+Ni-FeGe} \\
  & &  & &   & &  &  &  & $P6_{3}/mmc$  \cite{workon-non-cubic-Co+Ni-MnGe,workon-non-cubic-CoMnGe++Co+Ni-FeGe} \\
 CoMnP  &-100.06&-104.32  &-177.82&{\bf -187.28}  & 5.36 & 3.87, 5.14 &5.85, 3.45  &5.89, 3.46, 6.68  &$Pnma$ \cite{workon-non-cubic-Co-Cr+Mn+Fe-P+As++Ni-Cr+Mn+Fe-P+As,workon-non-cubic-Co+Ni-Cr+Mn+Fe-P++Co+Ni-Cr+Mn-As,workon-non-cubic-Co+Ni-Mn+Fe-P} \\
 CoMnAs &-57.81 & -31.45  & -74.94& {\bf -78.84}  & 5.54 & 4.05, 5.31 &6.12, 3.59  &6.31, 3.62, 6.97  &$Pnma$ \cite{workon-non-cubic-Co-Cr+Mn+Fe-P+As++Ni-Cr+Mn+Fe-P+As,workon-non-cubic-Co+Ni-Cr+Mn+Fe-P++Co+Ni-Cr+Mn-As,workon-non-cubic-CoMnAs} \\
 CoMnSb &{\bf -28.27} &   8.81  & 17.64 & 8.81    & 5.82 & 4.35, 5.43 &6.41, 4.00  &5.43, 3.35, 7.53  & $F\bar{4}3m$ \cite{workon-non-cubic-CoMnSb} \\
 CoMnS  & -1.42 & -18.71  &{\bf -55.40} & -52.94  & 5.44 & 3.65, 6.82 &5.79, 3.53  &5.98, 3.54, 6.74 &---   \\
 CoMnSe &{\bf -16.25} &  32.25  & -4.35 &-10.41   & 5.63 & 4.18, 5.45 &6.20, 3.67  &6.22, 3.80, 6.98  &--- \\
 CoMnTe &{\bf -22.52} &  55.71  & 46.83& --      & 5.86 & 4.42, 5.49 &6.48, 3.95  &--  &---   \\
 CoFeAl &  12.34& -70.16  &-54.54 & {\bf -83.53}  & 5.50 & 4.11, 4.92 &6.87, 2.66  &4.96, 3.95, 7.30  &-- \\
 CoFeGa &  31.36& -39.77  &-30.97 & {\bf -44.02}  & 5.52 & 4.12, 4.97 &6.89, 2.69  &5.00, 3.96, 7.33  &-- \\
 CoFeSi & -58.07&-104.96  &-106.88&{\bf -111.96}  &5.38 & 3.95, 4.19 &5.93, 3.42 &5.60, 3.61, 6.79  &$Pnma$ \cite{workon-non-cubic-Co+Ni-Cr+Mn+Fe-Si} \\
 CoFeGe & -16.63& -33.41  &-24.35 &{\bf  -34.15}  & 5.50 & 4.08, 5.50 &6.13, 3.48  &5.00, 3.92, 7.31 & $P6_{3}/mmc$ \cite{workon-non-cubic-CoMnGe++Co+Ni-FeGe} \\
 CoFeP  & -74.40& -78.98  &-158.53&{\bf -162.25}  & 5.36 & 3.86, 5.02 &5.72, 3.47  &5.69, 3.52, 6.52  &$Pnma$ \cite{workon-non-cubic-Co-Cr+Mn+Fe-P+As++Ni-Cr+Mn+Fe-P+As,workon-non-cubic-Co+Ni-Cr+Mn+Fe-P++Co+Ni-Cr+Mn-As} \\
 CoFeAs &-35.39 &  -6.29  & -43.08& {\bf -43.28}  & 5.54 &4.07, 5.13 &5.99, 3.60  &5.92, 3.67, 6.82  & $P\bar{6}2m$ \cite{workon-non-cubic-Co-Cr+Mn+Fe-P+As++Ni-Cr+Mn+Fe-P+As} \\
 CoFeSb & {\bf -7.16} &  34.19  & 55.79 & 34.32   & 5.81 & 4.33, 5.28 &6.37, 3.85  &5.27, 4.29,7.55 &---  \\
 CoFeS  & 15.59 &  68.60  &-24.59 & {\bf -28.08}  & 5.46 & 3.64, 5.79 &5.74, 3.60  &5.84, 3.64, 6.44  &-- \\
 CoFeTe &{\bf -18.52} &  86.20  & 80.25& --      & 5.86 & 4.40, 5.26 &6.32, 3.96  &-- &--   \\
\hline
\end{tabular} 
\end{table}

\begin{table}
Table 2. Formation energy and Total Magnetic Moment per formula unit for Ni$BC$, calculated for cubic, hexagonal and orthorhombic structures.   $GS_{expt}$ gives experimentally observed symmetries.\\
\begin{tabular}{|c|c|c|c|c|c|c|c|c|c|}
\hline Material   &$E_{F\bar{4}3m}$ & $E_{P6_{3}/mmc}$ & $E_{P\bar{6}2m}$ &$E_{Pnma}$ & $a$ &$a,c$ &$a,c$  &$a,b,c $ &$GS_{expt}$    \\
            &(kJ/mol)& (kJ/mol) & (kJ/mol) &(kJ/mol) &  (\AA)     & (\AA)        &(\AA)         &(\AA)        &         \\
            &        &          &          &         & $F\bar{4}3m$     & $P6_{3}/mmc$        &$P\bar{6}2m$         &$Pnma$        &         \\
\hline NiCrAl &  31.05& -53.58  &-11.08 & {\bf -54.66}  & 5.54 & 4.17, 5.01 &6.03, 3.79  &4.96, 4.13, 7.31 &--  \\
 NiCrGa &  42.03&{\bf -29.92}   & -1.91 & -29.80   & 5.53 &4.16, 5.08 &6.16, 3.69  &5.08, 4.16, 7.20 &-- \\
 NiCrSi & -31.86& -98.04  &{\bf -110.48}&-107.06  &5.44 & 3.98, 5.09 &5.88, 3.62  &5.75, 3.62, 6.89  &$Pnma$ \cite{workon-non-cubic-Co+Ni-Cr+Mn+Fe-Si} \\
 NiCrGe & 1.73  & -38.00  &{\bf -41.95} & -39.01  & 5.55 &4.08, 5.28 &6.07, 3.73  &5.89, 3.77, 7.16 &-- \\
 NiCrP  &-45.42 & -83.87  &-149.98&{\bf -153.25}  & 5.44 & 3.83, 5.47 &5.89, 3.51  &5.85, 3.53, 6.82 &$Pnma$ \cite{workon-non-cubic-Co-Cr+Mn+Fe-P+As++Ni-Cr+Mn+Fe-P+As,workon-non-cubic-Co+Ni-Cr+Mn+Fe-P++Co+Ni-Cr+Mn-As} \\
 NiCrAs &-18.77 & -16.16  & -63.09& {\bf -67.03}  & 5.61 & 4.03, 5.49 &6.21, 3.63 &6.14, 3.68, 7.10  & $P\bar{6}2m$ \cite{workon-non-cubic-Co-Cr+Mn+Fe-P+As++Ni-Cr+Mn+Fe-P+As,workon-non-cubic-Co+Ni-Cr+Mn+Fe-P++Co+Ni-Cr+Mn-As} \\
 NiCrSb & {\bf -8.08} &  20.64  & 13.22 & 51.76   & 5.89 & 4.32, 5.65 &6.56, 3.92  &5.24,4.35, 7.54  &-- \\
 NiCrS  & 53.56 & -2.16   &{\bf -30.50} & -21.01  & 5.57 & 3.70, 6.80 &5.89, 3.59  &5.46, 3.64, 6.92 &--\\
 NiMnAl & -9.60 & {\bf -97.46}  &-55.52 & -97.00  & 5.61 & 4.13, 5.13 &6.03, 3.78  &5.12, 4.05, 7.30  &---  \\
 NiMnGa &  1.73 & {\bf -73.04}  &-46.11 & -4.23   & 5.64 & 4.13, 5.18 &6.28, 3.54  &4.90, 3.99, 7.33  & $P6_{3}/mmc$ \cite{workon-non-cubic-NiMnGa} \\
 NiMnSi & -70.17&-122.69  &-134.95&{\bf -140.61}  & 5.45 & 3.95, 4.13 &5.98, 3.51  &5.85, 3.56, 6.89  &$Pnma$  \cite{workon-non-cubic-Co+Ni-Cr+Mn+Fe-Si} \\
 NiMnGe & -41.31& -64.70  & -71.22 & {\bf -73.12}  & 5.57 &4.08, 5.26 &6.18, 3.59  &6.01, 3.67, 7.11  &$Pnma$ \cite{workon-non-cubic-Co+Ni-MnGe,workon-non-cubic-NiMnGe} \\
        &       &         &        &       &  & &  &  & $P6_{3}/mmc$ \cite{workon-non-cubic-Co+Ni-MnGe,workon-non-cubic-NiMnGe} \\
 NiMnSn & -21.01& {\bf -23.29}  &  7.77 & --   & 5.89 & 4.38, 5.47 &6.38, 4.16  &--  &-- \\
 NiMnP  & -88.22& -94.71  &{\bf -162.77}&-162.51  & 5.46 & 3.73, 6.09 &5.85, 3.46  &5.86, 3.44, 6.76  &$Pnma$ \cite{workon-non-cubic-Co-Cr+Mn+Fe-P+As++Ni-Cr+Mn+Fe-P+As,workon-non-cubic-Co+Ni-Cr+Mn+Fe-P++Co+Ni-Cr+Mn-As} \\
   & &  & &  & &  &  &  &  $P\bar{6}2m$ \cite{workon-non-cubic-Co-Cr+Mn+Fe-P+As++Ni-Cr+Mn+Fe-P+As,workon-non-cubic-Co+Ni-Cr+Mn+Fe-P++Co+Ni-Cr+Mn-As} \\
 NiMnAs &-66.86 & -41.15  & -73.01& {\bf -71.68}  & 5.63 & 4.10, 5.48 &6.17, 3.68 &6.17, 3.77, 7.02  &$Pnma$ \cite{workon-non-cubic-Co-Cr+Mn+Fe-P+As++Ni-Cr+Mn+Fe-P+As,workon-non-cubic-Co+Ni-Cr+Mn+Fe-P++Co+Ni-Cr+Mn-As} \\
  & & & &  &  &  & &  & $P\bar{6}2m$ \cite{workon-non-cubic-Co-Cr+Mn+Fe-P+As++Ni-Cr+Mn+Fe-P+As,workon-non-cubic-Co+Ni-Cr+Mn+Fe-P++Co+Ni-Cr+Mn-As} \\
 NiMnSb &{\bf -58.71} &  -8.86  &  -1.33& -8.89   & 5.90 & 4.37, 5.56 &6.47, 4.03  &5.56, 4.37, 7.57  & $F\bar{4}3m$ \cite{workonNiMnSb} \\
 NiMnS  &-13.46 & -33.52  &{\bf -45.29} & -41.32  & 5.58 & 3.72, 6.98 &5.95, 3.65  &6.20, 3.60, 6.93  &-- \\
 NiMnSe &{\bf -33.25} &  -1.52  &-14.48 &-21.23   & 5.77 & 3.96, 6.91 &6.28, 3.82  &6.58, 3.79, 7.72  &--  \\
 NiMnTe &{\bf -30.35} &  31.85  &  77.43& --      & 6.01 & 4.49, 5.55 &6.52, 4.12  &-- &--   \\
 NiFeAl & -19.81 & -91.48  &-69.46 & {\bf -94.45}  & 5.55 & 4.09, 5.00 &6.95, 2.62  &4.96, 3.88, 7.49  &-- \\
 NiFeGa & -6.3  & {\bf -57.89}  &-44.79 & -12.21  & 5.56 & 4.11, 5.05 &6.97, 2.65  &4.94, 3.87,7.38  &-- \\
 NiFeSi & -34.59&-110.89  &-119.66&{\bf -122.06}  & 5.44 & 3.95, 4.95 &5.92, 3.42  &5.46, 3.63, 6.87  &$Pnma$ \cite{workon-non-cubic-Co+Ni-Cr+Mn+Fe-Si} \\
 NiFeGe & -8.06 & {\bf -45.75}  &-40.97 & -42.52  & 5.56 & 4.08, 5.10 &6.12, 3.50  &5.27,3.85, 7.21  & $P6_{3}/mmc$ \cite{JMMM38} \\
 NiFeP  & -30.26& -64.64  &{\bf -142.42}&-139.56  & 5.41 & 3.90, 5.06 &5.83, 3.36 &5.59,3.50,6.71  & $P\bar{6}2m$ \cite{workon-non-cubic-Co-Cr+Mn+Fe-P+As++Ni-Cr+Mn+Fe-P+As,workon-non-cubic-Co+Ni-Cr+Mn+Fe-P++Co+Ni-Cr+Mn-As} \\
 NiFeAs & -5.65 & -12.95  & {\bf -39.64}& -36.52  & 5.59 & 4.12, 5.14 &6.03, 3.62  &5.47, 3.70, 7.05  & $P\bar{6}2m$ \cite{workon-non-cubic-Co-Cr+Mn+Fe-P+As++Ni-Cr+Mn+Fe-P+As,workon-non-cubic-Co+Ni-Cr+Mn+Fe-P++Co+Ni-Cr+Mn-As} \\
 NiFeS  & 47.08 &  41.83  &-9.06  & {\bf -10.83}  & 5.51 & 4.06, 5.11 &5.84, 3.66  &5.64, 3.50, 7.21  &--  \\
\hline
\end{tabular} 
\end{table}

\begin{table}
Table 3. Formation energy and Total Magnetic Moment per formula unit for Pt$BC$, calculated for cubic and hexagonal structures.  $GS_{expt}$ gives experimentally observed symmetries.\\
\begin{tabular}{|c|c|c|c|c|c|c|c|c|c|}
\hline Material   &$E_{F\bar{4}3m}$ & $E_{P6_{3}/mmc}$ & $E_{P\bar{6}2m}$ &$E_{Pnma}$ & $a$ &$a,c$ &$a,c$  &$a,b,c $ &$GS_{expt}$    \\
            &(kJ/mol)& (kJ/mol) & (kJ/mol) &(kJ/mol) &  (\AA)     & (\AA)        &(\AA)         &(\AA)        &         \\
            &        &          &          &         & $F\bar{4}3m$     & $P6_{3}/mmc$        &$P\bar{6}2m$         &$Pnma$        &         \\
\hline PtCrAl & -60.54& {\bf -146.51} &-101.12&-111.22 & 5.85 & 4.34 , 5.50 & 6.61 , 3.70 &5.07, 4.14, 7.91  &-- \\
 PtCrGa & -22.20& {\bf -95.85}  &-63.42&-45.77 & 5.86 & 4.34, 5.59  &6.70 , 3.66 &5.16, 4.14, 7.20    &-- \\
 PtCrIn & -0.83 & {\bf -30.08}  & 0.88 &-30.06  & 6.29 & 4.61, 5.76 & 7.82, 2.99 &5.77, 4.63, 7.96  &--\\
 PtCrSi & -60.52&-115.22  & -122.29&{\bf-128.17} & 5.80 & 4.21, 5.46 &6.35, 3.77 &6.11, 3.88, 7.39 &--\\
 PtCrGe & -31.12&-62.25   &-58.21&{\bf-63.81} & 5.92 & 4.30, 5.80  & 6.55, 3.83&6.17, 3.99,7.65   &--\\
 PtCrSn & -36.02&{\bf -50.16}   &-8.03&{\bf-50.16}  & 6.24 & 4.58, 5.87 & 6.95, 4.02 &5.87, 4.58, 7.93  & $P6_{3}/mmc$  \cite{workon-non-cubic-Pt-Cr+Fe-Sn} \\
 PtCrP  & -28.22&-66.39   & -100.78&{\bf-102.87} & 5.84 & 3.98, 6.71 & 6.41, 3.66 &6.34, 3.72, 7.27 &-- \\
 PtCrAs & -27.42&-12.34   &{\bf -41.32}&-39.84 & 6.00 & 4.28, 6.19 & 6.75, 3.65 &6.37, 3.90, 7.65 &-- \\
 PtCrSb & {\bf -49.52}&-8.73 &-8.48&-8.65  & 6.22 & 4.60, 6.01 & 7.16, 3.74 &6.01, 4.62, 7.93  &-- \\
 PtCrTe & {\bf -9.81} &47.41    & 40.00&40.11 & 6.33 & 4.63, 6.20 & 7.42, 3.69 &6.07,5.19,7.41    &-- \\
 PtMnAl &-117.70&{\bf -184.86}  &-160.06 &-184.47 & 5.98 & 4.35, 5.41 & 7.29, 2.86 &5.33, 4.16, 7.84   & $P6_{3}/mmc$ \cite{JMMM38} \\
 PtMnGa &-83.55 &{\bf -131.60}  &-119.18&-128.34 & 6.00 & 4.36, 5.54 &7.34, 2.88 &5.41, 4.16, 7.89  & $P6_{3}/mmc$   \cite{JMMM38} \\
 PtMnIn &-54.71 &{\bf -70.14}   &-44.17&-70.11  & 6.27 & 4.62, 5.71 & 7.79, 2.97 &5.71, 4.64, 7.97 &-- \\
 PtMnSi &-109.35&-136.27  & -156.04&{\bf-159.28} & 5.82 & 4.21, 5.48 & 6.39, 3.65 &6.23, 3.75, 7.34 &-- \\
 PtMnGe &-85.23 &-84.27   & -88.04 &{\bf-88.20}  & 5.95 & 4.35, 5.60 & 6.61, 3.70 &6.27, 3.90, 7.59  & $P6_{3}/mmc$ \cite{workon-non-cubic-PtMnGe++PtCrSb} \\
 PtMnSn &{\bf -96.61} &-75.29   &-37.87&-75.22  & 6.22 & 4.61, 5.71 & 6.85, 4.18&5.71, 4.63, 7.96  & $F\bar{4}3m$  \cite{JMMM38} \\
 PtMnP  &-75.50 &-70.07   &{\bf -110.29}&-104.96 & 5.86 & 4.00, 6.78 &6.17, 3.63 &6.14, 3.56, 7.86  &--\\
 PtMnAs &{\bf -82.47} &-37.07   &-51.27&-54.10  & 6.02 & 4.41, 5.70 &6.68, 3.81 &6.36, 4.05, 7.58   &-- \\
 PtMnSb &{\bf -106.05}&-33.61  &-18.61&-33.71  & 6.23 & 4.64, 5.79 & 6.95, 4.08 &5.78, 4.67, 8.00   & $F\bar{4}3m$ \cite{JMMM38} \\
 PtMnSe &{\bf -32.21} &19.46    &7.34&--    & 6.14 & 4.52, 5.73 &7.21, 3.46  &-- \\
 PtMnTe &{\bf -56.86} &-   &33.40&7.32   & 6.35 & - & 7.44, 3.71 &5.90, 5.40, 7.40   &-- \\
 PtFeAl &-111.81& -160.19  &-133.87&{\bf-167.48} & 5.88 & 4.33, 5.24 &7.22, 2.76 &5.17, 4.05, 7.93   &-- \\
 PtFeGa &-71.15 &{\bf -96.88}   &-88.74 &-44.87  & 5.90 & 4.35, 5.31 & 7.31, 2.78 &5.24, 3.98, 7.82  &--\\
 PtFeIn &-23.01 & -23.72 &-3.53 & {\bf -23.77} & 6.16 & 4.60, 5.52 & 7.75, 2.87 &5.52, 4.63, 7.93  &-- \\
 PtFeSi &-71.51 &-23.72 & -122.73 & {\bf-126.21} & 5.89 & 4.60, 5.52 &6.07, 3.72 & 5.82, 3.85, 7.27 &-- \\
 PtFeGe &-44.45 &-45.35   &{\bf -46.14}&-21.62  & 5.91 & 4.33, 5.40 &6.30, 3.74 & 5.33, 3.95, 7.81  &-- \\
 PtFeSn &{\bf -48.64} &-31.42   &-2.19&-32.25   &6.16 & 4.58, 5.57 & 7.64, 2.99 &5.57, 4.71, 7.73  & $P6_{3}/mmc$  \cite{workon-non-cubic-Pt-Cr+Fe-Sn} \\
 PtFeP  &-9.32  &-14.53   &{\bf -104.74}&-87.24 & 5.81 & 4.26, 5.24 &6.05, 3.63& 5.95, 3.57, 7.75   &-- \\
 PtFeAs &-10.42 &12.40    &{\bf -25.70}&25.12  & 5.97 & 4.42, 5.40 & 6.32, 3.71 & 5.46, 3.92, 7.98  &-- \\
 PtFeSb &{\bf -26.60} &13.09    &17.47&8.64   & 6.18 & 4.62, 5.59 &6.58, 4.03 & 5.58, 5.05, 7.45   &-- \\
\hline
\end{tabular} 
\end{table}

\subsection{Analysis of Electronic Stability: Formation Energy}

We study Co, Ni and Pt-based half Heusler alloys, namely, 
 Co$BC$, Ni$BC$ and Pt$BC$ ($B$ = Cr, Mn and Fe; $C$ = Al, Si, P, S,  
Ga, Ge, As, Se, In, Sn, Sb and Te). We calculate the formation energy
of these alloys in different crystal symmetries, 
which include cubic $C1_{b}$ ($F\bar{4}$3m), orthorhombic ($Pnma$), 
as well as hexagonal ($P\bar{6}2m$ and $P6_{3}/mmc$) structures. 
As per the formation energy calculations, out of the total 108 
alloys, which have been studied in this work, 25 compounds are 
found to be energetically unstable in any of the symmetries 
probed here (with close to zero or positive value of $E_{form}$). 
35 materials have reasonably low absolute value of formation energy 
(below -50 kJ/mol per f.u.). Out of that, six compounds 
have too low a value of $E_{form}$ (lower than -10 kJ/mol per f.u.). 
It is to be noted here
 that a negative value of the formation energy obtained from the 
calculations indicates that at zero 
temperature, the compound is more stable than the bulk counterparts
of the constituent elements. With a low value of
 formation energy, the stability of the compound is expected to 
be less. 
It is observed that out of the 108 compounds the energetically 
unstable ones mostly have a $C$ atom which has a large atomic number 
($Z$), specifically for the Co and Ni-based alloys. 
Figure 1 depicts the optimized symmetry for each of the 
83 energetically stable compounds,
which is obtained on the basis of formation energy 
from our first-principles calculations. 
In this Figure, the symbols $o$, $c$, $h1$ and $h2$ signify 
$Pnma$, $F\bar{4}3m$, $P6_{3}/mmc$ and $P\bar{6}2m$ space-groups, 
respectively. From this figure, we observe that, for alloys having
Co as the $A$ atom, the lowest energy structure 
predominantly corresponds to the orthorhombic structure. Cubic 
symmetry is found to be the lowest energy state, primarily for cases, 
which have a high $Z$ element (Se, Sb and Te) as $C$ atom as well 
as Mn or Fe as $B$ atom. It is to be noted that 
for low $Z$ elements as $C$ atom, the lowest energy phase is,  
 without exception, either an orthorhombic structure 
or one of the hexagonal 
structures in all the three Co, Ni and Pt-based alloys. 
When $A$ atom is Ni, and $B$ atom is Mn, the alloys with 
a high atomic number element, namely, Se, Sb and Te as
 $C$ atom has cubic structure as the  
lowest energy state. However,
for other $C$ elements, it is observed that the $o$, $h1$ and
$h2$ symmetries are preferred.
For Pt-based systems, the situation is the same as 
in case of Co and Ni case: the cubic phase with high $Z$ elements, 
Se, Sn, Sb and Te as $C$ atom, has the lowest energy. However,
it is not the favored symmetry when the $C$ atom is having low $Z$.
Here we point out that there are two relevant databases in the 
literature, where symmetries 
of the lowest energy state of many of the materials studied here are
listed. We compare the results from these two databases here. 
Since Ref.\onlinecite{heuslershome} deals with only the
cubic $F\bar{4}3m$ symmetry for the half Heusler alloys, 
and also does not deal with the Pt-based
materials, we compare only the cubic symmetry cases. We see the
matching for the cases with $C$ atom with high atomic number $Z$, 
namely, Sn and Sb, which are expected to yield cubic ground state,
is good (Figure 1). 
Open Quantum Materials Database (OQMD)\cite{OQMD} is a more detailed 
database and goes beyond the Heusler alloy compounds. 
Out of our 108 cases, 23 systems are not listed in this database. 
65 materials have been listed there against the cubic symmetry 
(notably, except two, all the other Pt-based systems are listed to be
 having cubic ground state) and overall a reasonably good matching is
observed between our results and the data from this database. 

To understand the relative stability of various symmetries in 
different compounds, the formation energies of the energetically 
stable 83 HHAs are shown in Tables 1 to 3. 
From the formation energy values we find that, many 
of the 83 materials are likely to exist in more than one 
crystal symmetry since the formation energies of these different 
symmetries are within a few meV per formula unit (f.u.) of each other.
 In Tables 1, 2 and 3, we highlight (in bold) the entries             
corresponding to the lowest formation energy. The experimentally 
available structures are also presented in these tables against the  
respective compound. We find that the predicted symmetries with the 
lowest energy for different materials, by and large, match with the 
literature, except for a very few materials.

{\bf Results on Cubic case} -- 
From Figure 1, we observe that group VIA seems the most favorable 
atom for formation of the HHAs in the cubic symmetry with 
{\it $C1_{b}$} structure (space-group {\it $F\bar{4}3m$}). 
It is further observed that for larger $C$ atoms of other
groups as well (for example, In, Sn, and Sb) the cubic 
phase is more stable compared to other symmetries.
Among materials studied here, 68 compounds seem to have energetically
stable cubic phase, ground state or not. Out of that maximum (30) is
Pt-based compounds.
From theoretical study, many of these cubic 
compounds are reported to 
show half-metallic-like character in the cubic 
$C1_{b}$ structure.\cite{PRB95JM} However, only a few among these 
have been experimentally synthesized and found to possess cubic 
$C1_{b}$ structure.\cite{workon-non-cubic-CoMnSb,workonNiMnSb,JMMM38,workon-non-cubic-Pt-Cr+Fe-Sn} 

{\bf Results on Hexagonal $P6_{3}/mmc$ symmetry} -- 
In total 69 compounds out of 108 have negative formation energy  
 in the case of {\it hexagonal Ni$_{2}$In type} structure 
(with space-group {\it $P6_{3}/mmc$}). 
Compared to the Co atom, with Ni and Pt atoms at 
the $A$ site, more number of 
alloys seems to be having negative formation energy.
Experimentally, CoFeGe  
is reported to have hexagonal structure. However, from our 
calculations the lowest energy state of CoFeGe is found to be 
orthorhombic;
 but, the formation energy in the hexagonal ($P6_{3}/mmc$) is 
close to that of the orthorhombic structure (Table 1). 
CoMnSi, CoMnGe and NiMnSi compounds are reported to exist in 
both hexagonal ($P6_{3}/mmc$) and orthorhombic ($Pnma$) structure. 
Our present calculation shows that the lowest energy state structure 
of these compounds is orthorhombic. 
 
{\bf Results on Hexagonal $P\bar{6}2m$ symmetry} -- 
From our calculations, 68 compounds are found to possess a negative
formation energy for the 
 {\it hexagonal} structure with a space-group of {\it $P\bar{6}2m$}.   
Out of these compounds quite a few are experimentally synthesized and 
are indeed found to have hexagonal $P\bar{6}2m$ 
structure.\cite{workon-non-cubic-Co-Cr+Mn+Fe-P+As++Ni-Cr+Mn+Fe-P+As,workon-non-cubic-Co+Ni-Cr+Mn+Fe-P++Co+Ni-Cr+Mn-As,workon-non-cubic-Co+Ni-Mn+Fe-P} 
However, there are few small differences. For example, 
NiCrSi is experimentally observed to have an orthohrombic structure. 
But from our calculation the lowest energy state is found to be 
hexagonal $P\bar{6}2m$, though it is to be noted here 
that the formation energy of 
orthorhombic structure is very close to that of 
the hexagonal $P\bar{6}2m$ structure. The energy difference is 3.42 
kJ/mol (34.2 meV per f.u.) which is of the order of the 
thermal energy. Experimentally, NiMnGe is found to possess 
orthorhombic and hexagonal ($P6_{3}/mmc$) structures. But from 
our calculation, the obtained lowest energy state is the orthorhombic
structure. We see from Table 2 that, 
the formation energies for the orthorhombic and the two 
hexagonal structures are reasonably close to each other. 
Similar is the case for CoFeAs. It is seen to have a 
hexagonal $P\bar{6}2m$ structure from literature. But from our 
calculations, it is observed that both the $Pnma$ and the 
hexagonal phase have very close value of $E_{form}$, and the 
former is slightly more stable than the other (Table 1).

{\bf Results on Orthorhombic phase} -- 
From our calculations the compounds, which are found to be in 
{\it orthorhombic, NiTiSn type} structure with space-group 
{\it $Pnma$}, are primarily Co-based compounds. In total 70 
alloys have energetically stable $Pnma$ structure.
It is worth-noting that for most of the NiMn$C$ structures, 
with a $C$ atom with a low $Z$ value, the ground state 
is found to be energetically very close to the orthorhombic phase. 
The symmetry of compounds at the lowest energy state 
 matches with the reports in the literature, barring a few exceptions.
While CoCrAs, CoMnGe and NiCrAs are found to exhibit a hexagonal 
symmetry, our calculations yield an orthorhombic lowest energy state
for these alloys. However, it is to be noted that 
from Table 1 and 2, we observe that the formation
energies of both these structures are very close (difference
 being 3 to 5 kJ/mol per f.u.). Another exception is the case of 
 CoMnAs. It is found to possess orthorhombic phase as ground state 
 but from experiment it is found to have $P6_{3}/mmc$ space-group. 
The energy difference, however, in this case, is large 
between these two phases (Table 1).
For many compounds it is found that, the orthorhombic phase has 
a formation energy which is very close (within 5 meV per f.u.) to 
the $E_{form}$ of one of the hexagonal phases.
Hence, from our detailed analysis of $E_{form}$ values, it can be 
inferred that 
the orthorhombic structure is the most common symmetry among 
the Co-based materials. There are quite a few of the Ni and Pt-based 
materials for which also this structure has the lowest energy. As
discussed above, in many cases two or more phases are close in energy.
Hence, it is likely that samples of the non-ground state phases of 
these materials may actually form 
under certain experimental conditions.

\begin{table}
Table 4. Geometry Analysis: $d$ gives the bondlength in \AA. Unit of density is Mg/m$^{3}$. In 3rd column, "Y" signifies there is good matching with the ground state XRD pattern. \\ 
\begin{tabular} {|c|c|c|c|c|c|c|} \hline
Material & Symmetry & XRD & density & $d$($A$-$B$) & $d$($A$-$C$) & $d$($B$-$C$)  \\ \hline
PtCrSn  & $P6_{3}/mmc$   & -  & 11.41  &  3.02  & 2.64, 2.94 & 3.02  \\
 	& $Pnma$    & Y   & 11.41  & 3.02, 3.02, 3.03, 3.03 & 2.64, 2.64, 2.94, 2.94 & 3.02, 3.02, 3.02, 3.03   \\
	& $F\bar{4}3m$ & N   & 10.01 &2.70  & 2.70  & 3.12   \\
	& $P\bar{6}2m$ & N   & 10.83 &3.03, 3.24 & 2.63, 2.68 & 2.90, 2.90, 2.91  \\
PtMnIn  & $P6_{3}/mmc$  & - & 11.48    &3.03  & 2.67, 2.86  & 3.03  \\
        & $Pnma$        & Y  & 11.48 &3.00, 3.00, 3.04, 3.04 & 2.67, 2.67, 2.86, 2.86 & 3.02, 3.03, 3.03, 3.03  \\
        & $F\bar{4}3m$ & N  &9.81  &2.72 & 2.72 & 3.14   \\
        & $P\bar{6}2m$ & N  &11.66  &2.97, 3.02 & 2.71, 2.80 & 2.83, 2.89, 2.89  \\
PtFeIn 	& $Pnma$  & - & 12.00 & 2.96,2.96,3.01, 3.01 & 2.66,2.66, 2.76, 2.76 & 2.99, 2.99, 2.99, 3.00 \\
        & $P6_{3}/mmc$   & Y & 12.00   &  2.99  & 2.66, 2.76 & 2.99 \\
	& $F\bar{4}3m$ & N  &10.39  &2.67  & 2.67  & 3.08 \\
	& $P\bar{6}2m$ & N  &12.21  &2.95, 2.98 & 2.69, 2.78 & 2.79,2.87, 2.87 \\
PtMnSn & $F\bar{4}3m$ & -- & 10.17    &2.69   & 2.69  & 3.11  \\
       & $P6_{3}/mmc$  & N & 11.65   &3.02 & 2.66, 2.86 & 3.02 \\
       & $Pnma$   & N & 11.65 &3.00,3.00, 3.03, 3.03 & 2.66,2.66, 2.86, 2.86 & 3.02, 3.02, 3.02, 3.02  \\
       & $P\bar{6}2m$ & N & 10.80 &2.87 & 2.62, 2.70 & 2.89, 3.18 \\
PtMnSb  & $F\bar{4}3m$ & - &10.20  &2.70  & 2.70  & 3.12    \\
 	& $Pnma$    & N & 11.44 & 3.02,3.02, 3.06,3.06 & 2.68,2.68,2.89, 2.89 & 3.04,3.04, 3.05,3.05 \\
	& $P6_{3}/mmc$   & N & 11.44 &3.05 & 2.68,2.90 & 3.05    \\
	& $P\bar{6}2m$ & N &10.87  &3.04 & 2.61, 2.72 & 2.90,2.92, 2.92 \\
NiMnAs  & $P\bar{6}2m$& - & 7.75 &2.77, 2.88 & 2.35, 2.41 & 2.55, 2.62 \\
 	& $Pnma$   & N & 7.67&2.75, 2.85, 2.90, 2.91& 2.34, 2.36, 2.42,2.69 & 2.56, 2.60, 2.69 \\
	& $F\bar{4}3m$ & N &7.00  &2.44   & 2.44  &2.82 \\
	&$P6_{3}/mmc$   & N & 7.84 &2.74 & 2.37,2.74&2.74 \\
NiMnAl  & $P6_{3}/mmc$   & - & 6.16 &2.71 & 2.39, 2.57 & 2.71 \\
 	& $Pnma$  & N & 6.18 &2.53, 2.64, 2.78,2.93 & 2.41, 2.44, 2.54, 2.58 & 2.63, 2.64, 2.72, 2.95 \\
	& $P\bar{6}2m$ & N &    &     &     &       \\
	& $F\bar{4}3m$ & N & 5.26&2.43 & 2.43 & 2.81     \\
CoCrGe & $P6_{3}/mmc$   & - & 8.20 &2.69 & 2.37, 2.55 & 2.69 \\
       & $Pnma$  & N &7.85 &2.65, 2.77, 2.79, 2.81 & 2.33, 2.36, 2.37 & 2.58, 2.59, 2.63 \\
       & $P\bar{6}2m$ & N &7.87  &2.71, 2.90 &2.33, 2.36  & 2.57, 2.58 \\
       & $F\bar{4}3m$ & N   &7.37 &2.38  & 2.38  &  2.75 \\ \hline
\end{tabular} 
\end{table}

\subsection{Geometry Analysis: Lowest Energy State versus Other States}

The lattice parameters for all the energetically stable materials  
are given in Tables 1 to 3. 
The cubic phase has a relatively open structure. Hence, 
a cubic phase is found to be the lowest energy state for
materials with atoms with larger atomic radii and larger $Z$ 
values (of Pt and some of the $C$ atoms). 

There are about twelve systems where the $E_{form}$ values of two 
phases are very close and these values are within 1 kJ/mol per f.u.
To understand this small difference in the $E_{form}$, we 
perform a detailed geometry analysis.
Table 4 gives relevant data for some typical materials, where in a 
few cases there is an excellent matching of the $E_{form}$ between 
two symmetries and where there is no matching. In this table, 
along with the
density values and the bondlength between two atoms ($d$), we have 
also noted if the simulated X-ray diffraction (XRD) pattern\cite{VASP} 
of the phase matches with the same of the ground state structure
 or not. We present the data of all the symmetries for each material
where first, 
second, third and fourth rows correspond to the ground state ($GS$) 
and the three other phases, having difference in formation energy 
with the ground state in an increasing order: these are described 
as $GS$+1, $GS$+2 and $GS$+3 states. For PtCrSn 
and PtMnIn it is observed from Table 3 
that, $P6_{3}/mmc$ is the ground state. However, it is seen that the 
respective $Pnma$ phases, that is the $GS$+1 phase   
in these two systems, possess
 a $E_{form}$ energy which is very close to the ground state (within
0.01 kJ/mol/f.u.). Similarly, PtFeIn has a $Pnma$ structure as
the lowest energy state and $P6_{3}/mmc$ structure also has a very 
similar value for the formation energy (Table 3). From Table 4 it
is clear that the XRD patterns of these two symmetries for 
the three above-mentioned materials are expected to be close.   
 Additionally, the densities of these two phases for PtCrSn, PtMnIn 
and PtFeIn are close to each other.
 When we compare the bondlength $d$ between the atoms
$A$-$B$, $A$-$C$ and $B$-$C$ for the two symmetries for these three
materials, we note that the values vary
maximum by only $\pm$0.03 \AA. Since the density, bondlengths and the
simulated XRD patters match so well, it is clear that the internal
local geometries of each atom and subsequently the bonding nature
in the two phases with two different symmetries for each of these
three materials are the same. 
It is also clear from Tables 1 to 4 that
the other two symmetries ($F\bar{4}3m$ and $P\bar{6}2m$)
 are not only energetically farther from the ground state, but these 
are also different from the geometric point of view.
In case of NiCrGa (Table 2), and PtCrIn (Table 3) also
 similar results are obtained. While these two symmetries 
($P6_{3}/mmc$ and $Pnma$) in NiCrGa show good overall matching, 
in case of PtCrIn, the XRD patterns
are found to be not quite close.

Further, it is interesting to probe PtMnSn and PtMnSb for the 
following reason.
It is observed that the ground state symmetry is cubic in both the
cases. When the data for these two materials from Table 4 are analyzed
it is found that, while the geometric data of the ground state 
symmetry does not match with those of any of the other three 
symmetries, all the data from second and third row (for phases 
$Pnma$ and 
$P6_{3}/mmc$) match very well. Table 3 lists the respective 
formation energies for these two materials and we observe that 
these data are indeed consistent with this observation. 
Subsequently, the XRD patterns of 
these materials for the two above-mentioned symmetries are seen 
to resemble each other. 
Next we discuss a few cases, where the $E_{form}$ value of one 
symmetry is only somewhat close to the ground state. We take the 
example of NiMnAs. The ground state $P\bar{6}2m$ has a $E_{form}$
(-73.01 kJ/mol per f.u.) and density values (7.75 Mg/m$^{3}$); 
 the $Pnma$ symmetry has somewhat close values for these two
quantities (-71.68 kJ/mol per f.u.
and 7.67 Mg/m$^{3}$, respectively). Consequently, the geometric data
including simulated XRD patterns of the two phases also exhibit not 
a good matching with each other. Similar is the case for materials, 
for example, NiMnAl, CoCrGe, PtMnAl and PtMnGe.

\begin{table}
Table 5. Total and Partial Magnetic Moment per formula unit (in units of $\mu_{B}$) for Co$BC$, calculated for cubic phase ($CS$) and the lowest energy structure ($GS$). The numbers are put in italics when the corresponding formation energy is positive. NVE is the number of valence electrons, $SP$ is the spin polarization at the Fermi level. \\ 
\begin{tabular}{|c|c|c|c|c|c|c|c|c|c|c|c|}
\hline Material & NVE & $CS$ & & & & & $GS$ & & & &  \\\hline
& & $\mu_{T}$ &$\mu_{A}$ &$\mu_{B}$  &$\mu_{C}$ & $SP$ 
&  $\mu_{T}$ &$\mu_{A}$ &$\mu_{B}$  &$\mu_{C}$ & $SP$ \\\hline
 CoCrAl & 18 & {\it 0.00 }   & {\it 0.00}   & {\it 0.00}   & {\it 0.00} &{\it 0.43}
 & 1.43 (Pnma)  &0.37 & 1.02  &-0.01 & 54.5 \\
 CoCrGa & 18 &  {\it 0.00} &  {\it 0.00}  & {\it 0.00}   & {\it 0.00} &{\it 0.51} 
 &  1.57 (Pnma) &0.36 & 1.18  &-0.02 & 52.4 \\
 CoCrSi & 19 & 1.00 &-0.17 &1.17  &-0.03 & 99.9
 & 0.99 (Pnma) &-0.03 &1.04  &-0.02&   \\
 CoCrGe & 19 & 1.00 & -0.30&1.31  &-0.05 & 100
 & 2.40 ($P6_{3}/mmc$) & 0.55 &1.84  &-0.05  &7.03 \\
 CoCrP & 20 & 2.00 & -0.01&1.95  &-0.05 & 100
 & 1.93 (Pnma) & 0.24 &1.73  &-0.04 & 76.3  \\
 CoCrAs & 20 & 2.00 &-0.25 &2.19  &-0.07 & 100
 & 2.18 (Pnma)  &-0.15 &2.33  &-0.07 & 52.2 \\
 CoCrS & 21 & {\it 3.00 }  & {\it -0.17 }  &{\it 2.81 }   &{\it 0.02 }& {\it 97.9 }
 & 2.94 (Pnma) &0.29 &2.61  &-0.04 & 58.0 \\
 CoMnAl & 19 & {\it 1.03}   & {\it -0.20}  &{\it 1.36}  &{\it -0.03} & {\it 78.7} 
 & 3.32 (Pnma) &0.78 &2.65  &-0.04 & 18.4 \\
 CoMnGa & 19 & {\it 3.00 } & {\it 0.37 }  &{\it 2.69 }  &{\it -0.06 } & {\it 95.6 }
 & 4.12 (Pnma) &1.15 &3.12  &-0.09 & 34.7 \\
 CoMnSi & 20 & 2.00 &0.03 &2.14  &-0.06 & 100
 & 3.49 (Pnma) &0.67 &2.98  &-0.06 & 41.2 \\ 
 CoMnGe & 20 & 2.00 &-0.19&2.36  &-0.09 & 100
 & 3.75 (Pnma) &0.71  &3.22  &-0.09 & 44.0  \\
 CoMnP & 21 & 3.00 & 0.11 &2.90  &-0.06 & 100 
 & 2.99 (Pnma) &0.32  &2.79  &-0.05 & 13.8 \\
 CoMnAs & 21 & 3.00 &-0.06&3.08  &-0.04 & 100
 & 3.10 (Pnma) &0.12  &3.09  &-0.07 & 52.9  \\
 CoMnSb & 21 &      &      &      &      &
 & 3.00 (Cubic) & 0.15 &3.24  &-0.08 &100  \\
 CoMnS & 22 & 4.00 & 0.42 &3.37  &0.03 & 100
 & 2.39 ($P\bar{6}2m$) &-0.09 &2.52  &-0.06  & 13.8 \\
 CoMnSe & 22 &      &      &      &     & 
 & 4.00 (Cubic) & 0.37 &3.47  &0.00 &97.8  \\
 CoMnTe & 22 &      &      &      &      &
 & 4.00 (Cubic) & 0.38 &3.56  &-0.03 &100  \\
 CoFeAl & 20 & {\it 2.64 }&  {\it 0.64 }& {\it 2.27 }  & {\it -0.05 }& {\it 58.7}
 & 3.26 (Pnma) &1.03  &2.41  &-0.04 &49.7  \\
 CoFeGa & 20 & {\it 2.72 }& {\it  0.61 } &{\it  2.36}   & {\it -0.08 } &{\it  21.4 } 
 & 3.41 (Pnma) &1.04  &2.52  &-0.06 & 45.3 \\
 CoFeSi & 21 & 3.00 & 0.65 &2.52  &-0.05 & 94.4
 & 2.56 (Pnma) &0.56  &2.18  &-0.04 &66.0  \\
 CoFeGe & 21 & 3.00 & 0.51 &2.71  &-0.09 & 100
 & 2.72 (Pnma) &0.62  &2.25  &-0.06 &63.7  \\
 CoFeP  & 22 & 3.86 & 0.99 &2.84  &0.01 & 1.1
 & 2.04 (Pnma) &0.36  &1.80  &-0.04 &23.2  \\
 CoFeAs & 22 & 3.98 &1.04  & 2.93 & 0.0 & 66.1
 & 2.14 (Pnma) &0.25  &2.02  &-0.04 & 45.9 \\
 CoFeSb & 22 &      &      &      &      & 
 & 3.99 (Cubic) & 1.04 &2.99  &-0.02  &58.3 \\
 CoFeS  & 23 & {\it 4.94 } & {\it 1.55 } & {\it 3.13 } & {\it 0.19 } & {\it 18.7}
 & 2.98 (Pnma) &0.62  &2.39  &0.03 & 33.6  \\
 CoFeTe & 23 &      &      &      &     &
 & 4.42 (Cubic) & 1.25 &3.07  &0.07  &69.5 \\
\hline
\end{tabular} 
\end{table}

\begin{table}
Table 6. Total and Partial Magnetic Moment per formula unit (in units of $\mu_{B}$) for Ni$BC$, calculated for cubic ($CS$) and the lowest energy structure ($GS$). The numbers are put in italics when the corresponding formation energy is positive.  NVE is the number of valence electrons, $SP$ is the spin polarization at the Fermi level. \\ 
\begin{tabular}{|c|c|c|c|c|c|c|c|c|c|c|c|}
\hline Material & NVE & $CS$ & & & & & $GS$ & & & &  \\\hline
& & $\mu_{T}$ &$\mu_{A}$ &$\mu_{B}$  &$\mu_{C}$ & $SP$ 
&  $\mu_{T}$ &$\mu_{A}$ &$\mu_{B}$  &$\mu_{C}$ & $SP$ \\\hline
 NiCrAl & 19 & {\it 1.00 } & {\it -0.03 } &{\it 1.13 } &{\it -0.04 } &{\it 100}
 & 2.17 ($Pnma$) & 0.28 &1.78  &-0.01 & 2.8 \\
 NiCrGa & 19 &  {\it 1.00 } & {\it  -0.10 } &{\it  1.19 }  &{\it  -0.06 } &{\it 100}
 & 2.35 ($P6_{3}/mmc$) & 0.28 &2.00  &-0.03  &9.4 \\
 NiCrSi & 20 & 2.00 & 0.11 &1.93  &-0.07 & 100
 & 2.30 ($P\bar{6}2m$) & 0.23 &2.14  &-0.06  &4.4 \\
 NiCrGe & 20 & {\it  2.00 } & {\it  0.02}  & {\it 2.05 }  &{\it  -0.11} & {\it 100}
 & 2.79 ($P\bar{6}2m$) & 0.18 &2.66  &-0.07  &26.5 \\
 NiCrP  & 21 & 3.00 & 0.17 &2.69  &-0.08 & 100
 & 2.62 ($Pnma$) & 0.17 &2.47  &-0.06 &46.7  \\
 NiCrAs & 21 & 3.00 & 0.06 &2.84  &-0.11 & 100
 & 2.99 ($Pnma$) & 0.09 &2.87  &-0.09 & 69.6 \\
 NiCrSb & 21 &      &      &      &      & 
 & 3.05 (Cubic) & 0.01 &2.99  &-0.11 &81.6  \\
 NiCrS  & 22 & {\it 3.99 } & {\it 0.26 } & {\it 3.30 }  & {\it 0.01 } & {\it 98}
 & 2.40 ($P\bar{6}2m$) & 0.08 &2.24 &-0.06  &5.6 \\
 NiMnAl & 20 & 3.12 & 0.19 &3.15  &-0.07 & 76.8
 & 3.20 ($P6_{3}/mmc$) & 0.25 &2.99  &-0.04  &47 \\
 NiMnGa & 20 & {\it 3.32}  & {\it 0.16 } &{\it 3.33 }  & {\it -0.11 } & {\it 56.6}
 & 3.26 ($P6_{3}/mmc$) & 0.25 &3.10  &-0.07  & 38.5\\
 NiMnSi &  21 & 3.00 & 0.20 &2.96  &-0.08 & 100
 & 2.83 ($Pnma$) & 0.16 &2.81  &-0.06 & 57.1 \\
 NiMnGe & 21  & 3.01 & 0.10 &3.09  &-0.13 & 66.3
 & 2.98 ($Pnma$) & 0.09 &3.06  &-0.10 &  43.3 \\
 NiMnSn & 21  & 3.35 & 0.06 &3.47  &-0.12 & 1.3
 & 3.49 ($P6_{3}/mmc$) & 0.17 &3.40  &-0.07  &35.2  \\
 NiMnP  & 22 & 4.00 & 0.38 &3.51  &-0.03 &100
 & 2.33 ($P\bar{6}2m$) & 0.01 &2.42  &-0.08  & 48.4\\
 NiMnAs & 22 & 4.00 & 0.32 &3.61  &-0.05 &100
 & 3.44 ($P\bar{6}2m$) & 0.09 &3.41  &-0.05  & 8.9\\
 NiMnSb & 22 &      &      &      &      & 
 & 4.00 (Cubic) & 0.26 &3.72  &-0.06 & 100 \\
 NiMnS  & 23 & 4.99 & 0.65 &3.92  &0.16 & 88.8
 & 3.60 ($P\bar{6}2m$) & 0.20 &3.32  &0.00  &21.8 \\
 NiMnSe & 23 &      &      &      &     & 
 & 4.96 (Cubic) & 0.60 &4.00  &0.13 & 78.3 \\
 NiMnTe & 23 &      &      &      &     & 
 & 4.87 (Cubic) & 0.49 &4.06  &0.10 & 9.8 \\
 NiFeAl & 21 & 3.00 & 0.44 &2.76  &-0.04 &93.8
 & 2.48 ($Pnma$) & 0.32 &2.29  &-0.03  & 55.5\\
 NiFeGa & 21 & 3.00 & 0.41 &2.79  &-0.08 & 100
 & 2.64 ($P6_{3}/mmc$)  & 0.25 &2.53  &-0.05  &60.5 \\
 NiFeSi & 22 & 3.38 & 0.51 &2.86  &-0.01 & 76.8
 & 1.72 ($Pnma$) & 0.10 &1.76  &-0.04 & 71.2 \\
 NiFeGe & 22 & 3.56 & 0.54 &3.01  &-0.01 &72.4
 & 2.37 ($P6_{3}/mmc$) & 0.11 &2.38  &-0.05  &59.8 \\
 NiFeP  & 23 & 3.64 & 0.57 &2.94  &0.05 & 76.7
 & 1.08 ($P\bar{6}2m$) & 0.18 &0.96  &-0.03 &69.7 \\
 NiFeAs & 23 & 3.71 & 0.56 &3.03  &0.04 & 78.8
 & 2.08 ($P\bar{6}2m$) & -0.01&2.19  &-0.03  &57.2 \\
 NiFeS  & 24 & {\it 4.02 } & {\it 0.71 } &{\it 3.09 }  &{\it 0.15 } &{\it 97.2}
 & 2.00 ($Pnma$) & 0.13 &1.87  &0.01 & 21.6 \\
\hline
\end{tabular} 
\end{table}

\begin{table}
Table 7. Total and Partial Magnetic Moment per formula unit (in units of $\mu_{B}$) for Pt$BC$, calculated for cubic ($CS$)  and the lowest energy structure ($GS$). NVE is the number of valence electrons, $SP$ is the spin polarization at the Fermi level. \\ 
\begin{tabular}{|c|c|c|c|c|c|c|c|c|c|c|c|}
\hline Material & NVE & $CS$ & & & & & $GS$ & & & &  \\\hline
& & $\mu_{T}$ &$\mu_{A}$ &$\mu_{B}$  &$\mu_{C}$ & $SP$ 
&  $\mu_{T}$ &$\mu_{A}$ &$\mu_{B}$  &$\mu_{C}$ & $SP$ \\\hline
 PtCrAl & 19 & 1.00 & -0.07&1.17 & -0.04 &100
 & 3.23 ($P6_{3}/mmc$) & 0.09 &3.04 &-0.03 &30.4  \\
 PtCrGa & 19 & 1.00 & -0.09&1.18 &-0.06 &100
 & 3.28 ($P6_{3}/mmc$) & 0.06 &3.13 &-0.05 &36.8  \\
 PtCrIn & 19 & 3.97 & 0.07 &3.56 &0.0 &15.3
 & 3.68 ($P6_{3}/mmc$) & 0.06 &3.46 &-0.04 & 18.5 \\
 PtCrSi & 20 & 2.00 & -0.04&2.13 &-0.07 & 100
 & 3.08 ($Pnma$) & 0.07 &3.00 &-0.08 & 6.5 \\
 PtCrGe & 20 & 2.15 &-0.19 &2.43 &-0.14 &73.4
 & 3.56 ($Pnma$) & 0.06 &3.42 &-0.08 & 4.9  \\
 PtCrSn & 20 & 3.24 &-0.04 &3.25 &-0.09 & 20.9
 & 3.61 ($Pnma$) & 0.04 &3.47  &-0.07 & 42.7 \\
 PtCrP  & 21 & 3.00 &-0.05 &2.93 &-0.08 &100
 & 3.00 ($Pnma$) & 0.0  &3.03 &-0.09 & 66.4 \\
 PtCrAs & 21 & 3.04 &-0.09 &3.04 &-0.11 & 89.5
 & 3.16 ($P\bar{6}2m$) &-0.08 &3.26 &-0.13  &20.7 \\
 PtCrSb & 21 &      &      &     &      & 
 & 3.19 (Cubic) & -0.07&3.24 &-0.12 & 70.6 \\
 PtCrTe & 22 &      &      &      &      & 
 & 4.01 (Cubic) & 0.06 &3.66  &-0.06 & 95.5\\
 PtMnAl & 20 & 3.76 & 0.10 &3.73  &-0.05 & 70.7
 & 3.51 ($P6_{3}/mmc$) & 0.11 &3.47  &-0.04  &39.2 \\
 PtMnGa & 20 & 3.86 & 0.09 &3.81 &-0.07 & 61.8
 & 3.79 ($P6_{3}/mmc$) & 0.11 &3.73 &-0.06 & 35.3 \\
 PtMnIn & 20 & 4.14 & 0.10 &4.05 &-0.05 &53.5
 & 4.01 ($P6_{3}/mmc$) & 0.11 &3.93 &-0.05 & 18.5 \\
 PtMnSi & 21 & 3.05 & 0.0 &3.25 &-0.09 & 25.4
 & 3.05 ($Pnma$) & 0.0  &3.21 &-0.06 & 14.9 \\
 PtMnGe & 21 & 3.43 &-0.11 &3.70 &-0.17 &21.5
 & 3.26 ($Pnma$) & -0.02&3.45 &-0.10 &0.99  \\
 PtMnSn & 21 &      &      &      &      &
 & 3.67 (Cubic) & 0.02 &3.82 &-0.11 &  29.5\\
 PtMnP & 22 & 4.00 & 0.16 &3.76  &-0.04 & 100
 & 2.12 ($P\bar{6}2m$) & 0.04 &2.16 & -0.09& 84.2 \\
 PtMnAs & 22 &      &      &     &      &
 & 4.01 (Cubic) & 0.13 &3.85 &-0.08 &76.5 \\
 PtMnSb & 22 &      &      &      &      &
 & 4.02 (Cubic) & 0.11 &3.93 &-0.07 &61.9 \\
 PtMnSe & 23 &      &      &      &     &
 & 4.74 (Cubic) & 0.27 &4.19 &0.06 & 25.7\\
 PtMnTe & 23 &      &      &      &     &
 & 4.83 (Cubic) & 0.27 &4.26 &0.08 & 8.63\\
 PtFeAl & 21 & 3.00 & 0.19 &2.99  &-0.04 & 94.2
 & 2.55 ($Pnma$) & 0.12 &2.55 &-0.03  &64.8 \\
 PtFeGa & 21 & 3.00 & 0.17 &3.01  &-0.07 &100
 & 2.87 ($P6_{3}/mmc$) & 0.11 &2.88 &-0.04  &60.5 \\
 PtFeIn & 21 & 3.00 & 0.14 &3.10 &-0.07 & 100
 &3.02 ($Pnma$) & 0.11 &3.02 &-0.04  &64.2 \\
 PtFeSi & 22 & 3.30 & 0.27 &3.03 &-0.01 & 66.6
 & 1.92 ($Pnma$) & 0.02 &2.07 &-0.04  &46.5 \\
 PtFeGe & 22 & 3.94 & 0.38 &3.47 &0.01 &38.4
 & 1.99 ($P\bar{6}2m$) & 0.05 &2.06 &-0.06 &71.9  \\
 PtFeSn & 22 &      &      &     &      &
 & 3.53 (Cubic) & 0.26 &3.28 &-0.01 & 77.6 \\
 PtFeP  & 23 & 3.73 & 0.39 &3.16  &0.06 & 45.4
 & 1.18 ($P\bar{6}2m$) & 0.05 &1.24 &-0.05 &59.3 \\
 PtFeAs & 23 & 3.69 & 0.35 &3.21  &0.04 &62.3
 & 1.27 ($P\bar{6}2m$)  &0.0   &0.99 &-0.02  & 61.6 \\
 PtFeSb & 23 &      &      &      &     &
 & 3.57 (Cubic) & 0.28 &3.19 &0.02 & 65.1 \\
\hline
\end{tabular} 
\end{table}

\subsection{Analysis of Total and Partial Moment} 

In this subsection, we discuss the results on the magnetic properties
of the materials, studied here.
The calculations are carried out with a magnetic configuration
for all the materials. Since Co has a significant moment, there is a 
possibility that a ferrimagnetic (moments of Co and $B$ atom aligned 
anti-parallel to each other) configuration may be likely. 
After convergence we get, in a few cases a ferrimagnetic and in 
most of the cases a FM configuration as observed in the
literature.\cite{PRB95JM,PRB74VK} 
We present in Tables 5 to 7, the total, partial moments, the total
number of valence electrons, and the $SP$ 
 of all the materials, wherever possible.
The results of the cubic case and the lowest energy state obtained 
from our calculations are listed in these tables. When cubic is the
symmetry for the lowest energy state, the explicit entries 
 corresponding to the cubic case (on the left side of the Tables 
 5 to 7) are left empty. Values corresponding to energetically 
unstable cases have also been put (in italics) to see whether any 
trends which are found for the stable cases are followed by these 
or not.

{\bf Slater-Pauling rule, and Integer Moment versus Half-metallicity in Cubic case} -- 
It is observed in the literature that many Co-based Heusler alloys, 
specifically the half-metallic ones, follow the 
Slater-Pauling rule.\cite{JMMM423TR,JPD40HCK,PRB66IG} 
As a consequence of this 
rule, an almost linear variation of the magnetic moment with 
the atomic number of the $B$ atoms for the cubic case of the 
Co-based FHA materials is observed. In this work also we expect that 
a linearly increasing trend of the total moment as a function of the
$Z$ value of the $C$ atom will be observed in the cubic cases of 
Co$BC$. Figure 2 shows the CoMn$C$ cases. The cases with positive 
$E_{form}$ are also plotted for overall comparison.
We find that the linear trend is not quite followed when
the $E_{form}$ of the compound is positive. 
However, the total moment of the energetically stable alloys 
shows this linear trend as is clear from both Figure 2 and Table 5. 
These findings are true for CoCr$C$ systems as well.
We further observe from Figure 2 that though like the total moment, 
 the partial moments, mainly that of $B$ atom
 also show an increasing trend, there is a change
of slope in both the cases of the $A$ and $B$ atoms, as is true for 
 stable CoCr$C$ systems also. It is to be noted that 
none of these trends are followed in case of CoFe$C$.  
Figure 2 exhibits a few Ni and all Pt-based cases as well. 
In case of Ni and Pt-based systems, in general, none of these 
above-mentioned trends is observed. 
The moments of the NiFe$C$ alloys 
are seen to all together deviate from the trend. On the contrary, 
NiCr$C$ and NiMn$C$ cases with $C$ atoms 
from group IVA, VA and VIA tend to follow the Slater-Pauling rule,
when the $E_{form}$ is negative and $C$ atom has a lower
 atomic number. 
 As the group of the $C$ atoms changes, from IVA to VA to VIA, the 
total moment increases (by value 1), and partial moment of both 
Ni and $B$ atoms increases.
This observation has been discussed again in the next subsection 
in terms of the DOS of the up and down spin electrons.
As the group of the $C$ atom remains the same 
and the period changes, the total moment
remains the same, but moment of Ni and $B$ atoms decreases and
increases, respectively. 
For the Pt-based systems, the Slater-Pauling rule of linear 
increase of total moments is not obeyed (Figure 2).
Further, it has been observed that for the cubic cases, in many of 
the materials (maximum for the Co-based systems), the moment is 
integral in nature. It is expected that typically an integral 
moment leads to a half-metallic system in cubic phase. It will be 
discussed in detail in the next subsection after analyzing the 
total density of states of the up and down spin electrons, for
both cubic and non-cubic systems.  

{\bf Partial Moments in the Cubic case} -- 
Here we concentrate on the cubic phases of the materials.
It is observed from the partial moments that for some of 
the cases the final configuration has turned out to be ferrimagnetic.
 In the literature, it has been 
discussed\cite{PRB95JM} that there is a dependence of 
the long-range magnetic configuration on the number
of valence electrons (NVE). 
While systems with NVE=18  (case {\bf a}) show anti-ferromagnetism, 
cases with NVE=19 and 20 (case {\bf b})  exhibit ferrimagnetism. 
On the contrary, NVE=21 and 22 (case {\bf c}) lead to ferromagnetism. 
The $B$ atom carries the maximum moment in all these cases discussed 
here. The moment on the $A$ atom is almost equal in magnitude 
and opposite in direction in case {\bf a}. In case {\bf b}, moment
on $A$ is smaller in magnitude but oriented in an anti-parallel
arrangement with respect to moment of $B$ atom. 
On the contrary, case {\bf c} deals with a long range FM
configuration, where both $A$ and $B$ moments, though may be unequal, 
orient along the same direction. By analyzing our results (Tables 
5, 6 and 7) on the partial moments of the energetically
stable cubic phase, we find that there are only a few exceptions. 
While Co and Ni-based systems generally follow 
the trend, maximum deviation is observed in case of Pt-based systems.
We have found that for NVE=23 also, the results follow the trend as
in case {\bf c}. 

{\bf Total and Partial moments in the Ground states} --  
Here we concentrate on the total and partial moments of the 
phases with the lowest energy. Few of the cases exhibit
a close-to-integral total moment (CoMnP, CoFeP, CoFeS, NiCrAs, NiFeS,
PtCrP, PtMnSi, PtFeIn all in $Pnma$ symmetry as well as PtMnIn  
 in $P6_{3}/mmc$ symmetry). Like a cubic case, there may be a 
possibility of half-metallicity in these non-cubic cases. 
A detailed analysis of DOS is warranted for the 
validation of the same (see next subsection). 
We  analyze here the partial moments on the different atoms, 
present in the system. It is
seen that in most of the cases the moments on the $B$ and $C$ atoms
are anti-parellel to each other (Tables 5 to 7). 
However, the moments on the latter atoms are
 much smaller compared to the earlier ones as is observed 
in the cubic phase as well. Unlike the cubic case, no 
increasing trend in total moment as a function of $Z$ of the $C$ 
atom is observed for the lowest energy state. No trend is observed
 in the values of the total moments when the cubic and the 
$GS$ states of any of the materials is compared. In majority of the 
cases the moments on the $B$ and $A$ atoms are found to be parallel 
to each other, including the Co-based compounds. 

{\bf Spin Polarization at the Fermi Level} --  Next we
discuss about the extent of spin polarization at the Fermi level 
($SP$) of various 
materials in cubic versus the ground state. It is observed that 
except few of the $B$=Fe atom cases, Co$BC$ materials possess 
high SP in the cubic case.
For Ni$BC$, only with the exception of NiMnSn and NiMnTe, all the 
cubic cases exhibit high SP. 
On the other hand, for Pt$BC$, $SP$ for the cubic cases seems to
be below 50\% for many of the alloys. Many, but not all, of the 
cubic structures of different materials exhibit a 100\% $SP$. However,
this is not the case with the lowest energy structure of any
of the materials, which are studied in this paper.
In the lowest energy case, we observe that only a few materials, with
or without integral total moment, possess high $SP$, which is above
65\%. While there are 6 of these, but none has a 100\% $SP$. 
Out of these 6 cases, in ground state NiCrAs ($Pnma$) with a 
$SP$ of 69.6 \% and PtCrP ($Pnma$) with a $SP$ of 66.4 \% have a total 
integral moment of 2.99 and 3 $\mu_{B}$, respectively. 
Further, for PtFeGe ($P\bar{6}2m$) has a moment of 1.99 and
$SP$ value of 71.9 \%.
However, while in the ground state, CoCrP ($Pnma$), NiFeSi ($Pnma$), 
NiFeP ($P\bar{6}2m$), and PtMnP ($P\bar{6}2m$) have 
comparable $SP$ values of 76.3 \%, 71.2 \%, 69.7 \% 
and 84.2 \%, respectively, the corresponding total moments are 
1.93, 1.72, 1.08 and 2.12 $\mu_{B}$, respectively. 

\subsection{Analysis of Electronic Structure} 

{\bf Density of States of the Cubic Phase} -- 
For overall comparison of the trends, we plot in Figure 3, 4 and 5
 the up and down DOS of the cubic and the lowest energy state
of all the materials, including the ones with positive $E_{form}$.
It is observed that the trends which are followed in case of the
energetically stable ones are not strictly followed in case of
the ones with positive $E_{form}$. 
The following trend is observed from Figure 3 for Co$BC$ alloys. 
As the number of valence electrons of the $C$ atom 
increases in a period, for example, Al to Si to P to S, the valence 
 band width (VBW) is seen to increase systematically. Further,
specifically for all the other cases, except a $C$ atom from
 group VIA, the
DOS shifts away from Fermi level, leading to a higher binding energy
of the system. 
The DOS at and very close to Fermi level also changes.
In case of S, Se and Te, these systematics are not 
consistently followed throughout. 
Furthermore, it is seen that when the $C$ atom is from group IIIA, 
all the alloys of type Co$BC$ are found to be energetically unstable.
As the $Z$ value of $C$ atom increases in a group, for other groups,
for example, Si to Ge to Sn, {\it i.e.} atoms with same NVE, not only 
the VBW, but also the DOS at the Fermi level remains similar. 
When the $A$ atom changes, namely, Co atom is replaced by Ni 
and Pt, the shifting of the weight of the DOS curve towards 
lower energy (more binding energy) is clearly evident from 
Figures 4 and 5. The overall trends, near the Fermi level, as 
discussed above, are 
found to be the same. However, it is observed that while for 
$A$=Ni, most of the alloys are predicted to be energetically 
unstable, when the $A$ atom is from group IIIA. For Pt all the 
alloys seem to have negative $E_{form}$ and it is further to be
noted that the weight of the DOS shifts towards lower energy
making the systems more bound in comparison to $A$ = Co or Ni, as is 
evident from the formation energies given in Tables 1 to 3 as well. 
As to why there is dip in the total moment (Figure 2), when the $B$ 
atom is mainly an Mn atom and $C$ atom is from group IVA, can be 
understood by analyzing the respective DOS curves. The up and down 
DOS in this case are seen to be more compensated leading to a lower
moment when compared to the 
cases where $C$ atoms are from other groups (Figures 3 to 5). 
As the $Z$ value of $B$ atom increases and the $A$ and $C$ atoms
remain the same, the energy is lowered as $Z$ of $B$ atom
increases, since the NVE of the system increases. However, it has 
been noted that, largely, the down spin DOS is more involved in 
all these, compared to the DOS of the up spin.

{\bf Comparison of DOS of Cubic versus Lowest Energy State} -- 
Figure 6, 7 and 
8 give the total and partial DOS of the cubic and ground state
of few materials. We have chosen different materials with various
symmetries: while Figure 6 and 7 exhibit the DOS of $P6_{3}/mmc$
and $P\bar {6}2m$ symmetries, respectively, DOS of few materials 
for which $Pnma$ space-group is energetically the most favorable,  
is given in Figure 8. It is observed that as in these cases, 
and also in many other cases, there
is a shift of weight of DOS of the ground state towards lower 
energy compared to the cubic state. Further, after critical
analysis of DOS of ground versus the cubic state, the following few 
 general points become note-worthy and relevant with regard to the 
non-cubic ground state symmetry of many materials (Tables 1 to 3).
Mainly in the down spin DOS, a significant hybridization is noted 
between the $A$ and $C$ atoms, specifically, very close to 
the Fermi level.\cite{MBunpubl}
A double-peak structure between the DOS of these two atoms is
prominent in most of the cases. 
Further, hybridization between the $B$ and $C$ atoms and sometimes 
among all the three atoms is also observed. More than often 
this is found in the lower energy ranges (away from Fermi level) and
also typically for up spin DOS. While cases with both Mn and Cr 
as $B$ atoms show similar results, cases with Fe as $B$ atom are
slightly different: for example, the overlapping
behavior with $C$ and/or $A$ atoms is better for up spin DOS and 
at lower energy compared to the vicinity of the Fermi level. 
These observations may explain the following: 
 though, in the literature there is a lot of 
theoretical study on the cubic phases and half-metallicity in case 
of HHAs, experimentally many HHAs have been shown to prefer a lower 
symmetric structure and discussion about the HM-like behavior 
in a non-cubic case is generally missing from the literature. 

{\bf DOS of Two Symmetries with Close $E_{form}$} -- Figure 9
gives the density of states for a few materials, for which
two symmetries yield close to very close $E_{form}$ and also
geometry (Tables 1 to 4). As is evident from Table 3, for the
material PtCrSn, two symetries ($Pnma$ and $P6_{3}/mmc$) possess
the same $E_{form}$ within our calculational accuracy. The lowest 
panel in Figure 9 gives the support for the same from the electronic
structure calculations. The total as well as partial DOS for
 this material in these two symmetries have an excellent matching. 
Similar is the case for PtMnIn and PtFeIn where the $E_{form}$ values 
for these two phases are also very close.\cite{MBunpubl} 
 Further, two materials PtMnSn and PtMnSb have a cubic phase
as a lowest energy state. However, in the $GS$+1 (symmetry with
higher energy than the ground state) and $GS$+2 (symmetry with 
higher energy than the $GS$+1 state) cases, PtMnSn has $P6_{3}/mmc$
and $Pnma$, respectively, while PtMnSb has $Pnma$ and $P6_{3}/mmc$,
respectively. These two phases also have very similar formation 
energy. In Figure 9, we show the total and partial DOS for 
these $GS$+1 and $GS$+2 cases for PtMnSn and we observe that the 
DOS for these two cases are again matching very well.
Similar is the case for PtMnSb.\cite{MBunpubl} As is evident
from Tables 1 to 4 and Figure 9, the results of 
structural and electronic structure calculations are consistent
for all these five materials mentioned here.
On the contrary, when the $E_{form}$ for one phase is slightly
higer than the other (as in case of CeFeAs and NiMnAl), the peak 
positions and intensities of different peaks are not so alike as 
discussed above. Though in these two cases
the $E_{form}$ is different by only about 0.5 kJ/mol, from upper
two panels it is seen that the matching of DOS of the two phases 
is not so excellent. The matching is slightly better in
case of NiMnAl where the two symmetries involved are $P6_{3}/mmc$
and $Pnma$. This has been a general observation throughout. If 
$P6_{3}/mmc$ symmetry is found to be close to $Pnma$ symmetry,
the matching of the geometry and electronic structure are very 
good. However, regarding the matching between $Pnma$ and 
$P\bar{6}2m$ symmetries, 
geometrical and electronic structures are seen to match only 
reasonably for these two phases even 
if energetically these phases are close to each other.\cite{MBunpubl}
From literature on experimental studies of HHAs, it is observed 
that a very few cases exist where a material is reported to be 
in two different symmetries. Pertinent examples of such cases 
are NiMnAs, NiMnP, CoMnGe and NiMnGe. Figure 10 shows the DOS of 
the two experimentally reported phases for these materials. 
It is observed that both partial and total DOS are quite close 
for these two phases for all these materials. Since the difference
between the two $E_{form}$ values are more in case of CoMnGe, 
somewhat significant differences in total and partial DOS are 
visible from Figure 10.

{\bf DOS of Non-cubic HM-like States} -- Finally, we show the plot
of density of states for a few materials which are likely to exhibit
a HM-like behavior depending on the values of total (integral) moment  
in their non-cubic ground state. Figure 11 exhibits
the DOS of CoMnP, CoFeP, NiCrAs and PtCrP which show 
a gap or a pseudo-gap with a very low DOS at the Fermi level, 
for one of the spin channels, for both the ground state and the
cubic phase. The cases of NiCrAs and PtCrP in the lowest energy 
state ($Pnma$ phase) are reasonably
clear. A very small density of states at the Fermi level 
is observed for the down spin channel of both the materials. 
However, the DOS at the up spin channel is also
somewhat small for both the cases. This results in an effective spin 
polarization (at the Fermi level) of about 70 and 66\% for NiCrAs
and PtCrP in the ground state, respectively. 
Contrary to these cases, both CoMnP and CoFeP will probably behave
as a bad semi-metal rather than a half-metal since there are
small densities of states at both the spin channels, up DOS being
slightly larger in intensity than the down DOS.
We note from Tables 5 to 7 that, these four materials possess total 
moments which are very close to integers. Total moment of 3, 2, 2 and 
3 $\mu{B}$ are observed for CoMnP, CoFeP, NiCrAs and PtCrP, 
respectively. 
There are few other materials, for which also, in the non-cubic
case, the total moment is very close to an integer. These cases
include NiFeS, PtMnSi, PtFeIn - having the $Pnma$ (See Tables 5 to 7), 
and also PtMnIn - having the $P6_{3}/mmc$ space-group, respectively.
In these cases also, our calculations reveal the appearence of a 
pseudo-gap at the down spin channel.\cite{MBunpubl}  
However, the DOS at the Fermi level for the up spin channel, though 
is higher compared to the DOS for the down spin electron, the 
absolute value of the DOS for down spin electron is not negligible, 
as is observed in the case of NiCrAs and PtCrP. Further, 
we consider the cases of some other materials with high $SP$ but
with total moment not so close to an integer value. It has been 
observed that out of these materials, 
in the ground state, CoCrP and NiFeSi have $Pnma$, 
and NiFeP and PtMnP have $P\bar{6}2m$ symmetries; and as discussed 
above, these have total moment of 1.93, 1.72, 1.08 and 2.12 
$\mu_{B}$, respectively. The corresponding
$SP$ has been observed to be high and comparable to those of
 NiCrAs and PtCrP. Therefore, we observe that, though 
 there is no rigorous one-to-one relationship 
among the cubic symmetry, integral total moment and high $SP$,  
but a strictly half-metallic behavior (with exactly 100\% $SP$) 
and integral total moment are found to be associated only with the 
cubic symmetry. 
Hence, by combined analysis of magnetic and electronic structure 
calculations, CoCrP, NiCrAs, NiFeSi, NiFeP, PtCrP and PtMnP 
alloys are predicted 
to be non-cubic half Heusler alloys with significantly
high $SP$. However, the point to be noted here is that a high $SP$ 
does not necessarily indicate a possibility of semiconducting
behavior along one spin channel, and in turn, a possible application 
of a material as an appropriate spin-injector material. Hence
 we predict from our present density of states calculations that 
NiCrAs and PtCrP are the only two non-cubic materials which in 
their $Pnma$ phase may have a potential in this regard and also 
these are magnetic in nature. 
This observation awaits the experimental validation.         

\section{Conclusion}

In this paper geometric, electronic, and magnetic
properties of Ni, Co and Pt-based half Heusler alloys, namely, 
Ni$BC$, Co$BC$ and Pt$BC$ ($B$ = Cr, Mn and Fe; $C$ = Al, Si, P, S,  
Ga, Ge, As, Se, In, Sn, Sb and Te) have been calculated in detail 
using first principles calculations based on density functional 
theory. Quite a few of these materials with a $C$ atom from group
IIIA, IVA and VA have already been experimentally
 and/or theoretically found in various different symmetries. 
In this work, we probe the stability of all the above-mentioned 
alloys in different crystal symmetries, reported in the literature. 
These structures include, 
the most common (face-centered) cubic $C1_{b}$ phase (space-group 
$F\bar{4}3m$), and also orthorhombic (space-group $Pnma$), as well 
as hexagonal (space-groups $P\bar{6}2m$ and $P6_{3}/mmc$) phases. We 
find from our calculations of formation energy that along with alloys 
with $C$ elements from group IIIA, IVA and VA, alloys with $C$ 
elements from group VIA are also,  by and large, energetically 
stable. It has also been observed that high $Z$ elements as the $C$ 
atom lead to stabilized phases in case of the Pt-based compounds. On 
the contrary, it is not so in the case of Co and Ni-based materials. 

In literature
half-metallicity in many half and full Heusler alloys have been shown
to exist which is typically associated with a cubic symmetry. 
We note from the results of the magnetic properties calculations, 
that there is a possibility of existence 
of some novel non-cubic half-metallic-like half Heusler alloys, 
as these possess total integer moments. Therefore, 
 to discuss the relative stabilities of different symmetries  
in order to search for the respective lowest energy state for 
all the materials as well as to ascertain whether a material is
half-metallic or not, we analyze the partial and total density of 
states. Based on the results of the magnetic and electronic 
properties, (i) we show that for a material 
 depending on the hybridization between different atoms a particular 
 symmetry is more stable compared to the cubic or other phases; 
(ii) we observe that there is no rigorous {\it one-to-one 
relationship} between the {\it cubic symmetry} and 
{\it high spin polarization at the Fermi level};
(iii) it is found that a strictly half-metallic behavior (with 
100\% spin polarization) is associated only with the cubic symmetry; 
(iv) along with a few new cubic half-metallic alloys, we 
predict the possibility of existence of a few novel 
{\it non-cubic} alloys with significantly low DOS in one
 of the spin channels and {\it high spin polarization 
 at the Fermi level}. 

\section{Acknowledgement}

Authors thank P. A. Naik, T. Ganguli and Arup Banerjee for facilities 
and constant encouragement throughout the work. The scientific 
computing group of computer centre, RRCAT, Indore and P. Thander 
are thanked for help in installing and support in running the codes.

{}

\clearpage
\pagebreak

\begin{figure}
\centering
\includegraphics[width=0.8 \textwidth]{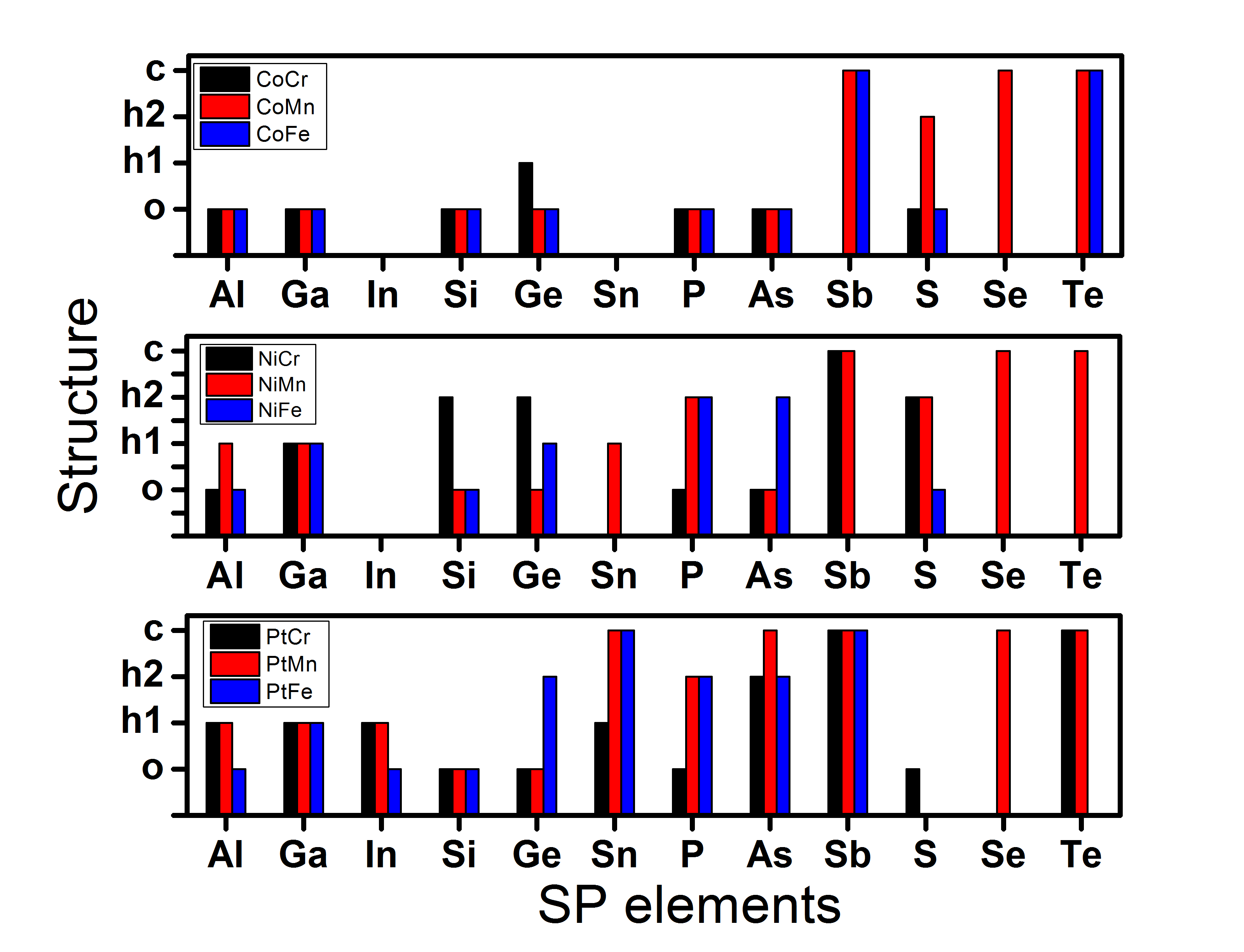}
\caption{
The optimized symmetry for each of the 83 energetically stable 
compounds, which is obtained on the basis of formation energy. 
$o$, $h1$, $h2$ and $c$ signify $Pnma$, $P6_{3}/mmc$, $P\bar{6}2m$
and $F\bar{4}3m$ space-groups, respectively.
}
\label{fig:1}
\end{figure}

\begin{figure}
\centering
\includegraphics[width=1.0 \textwidth]{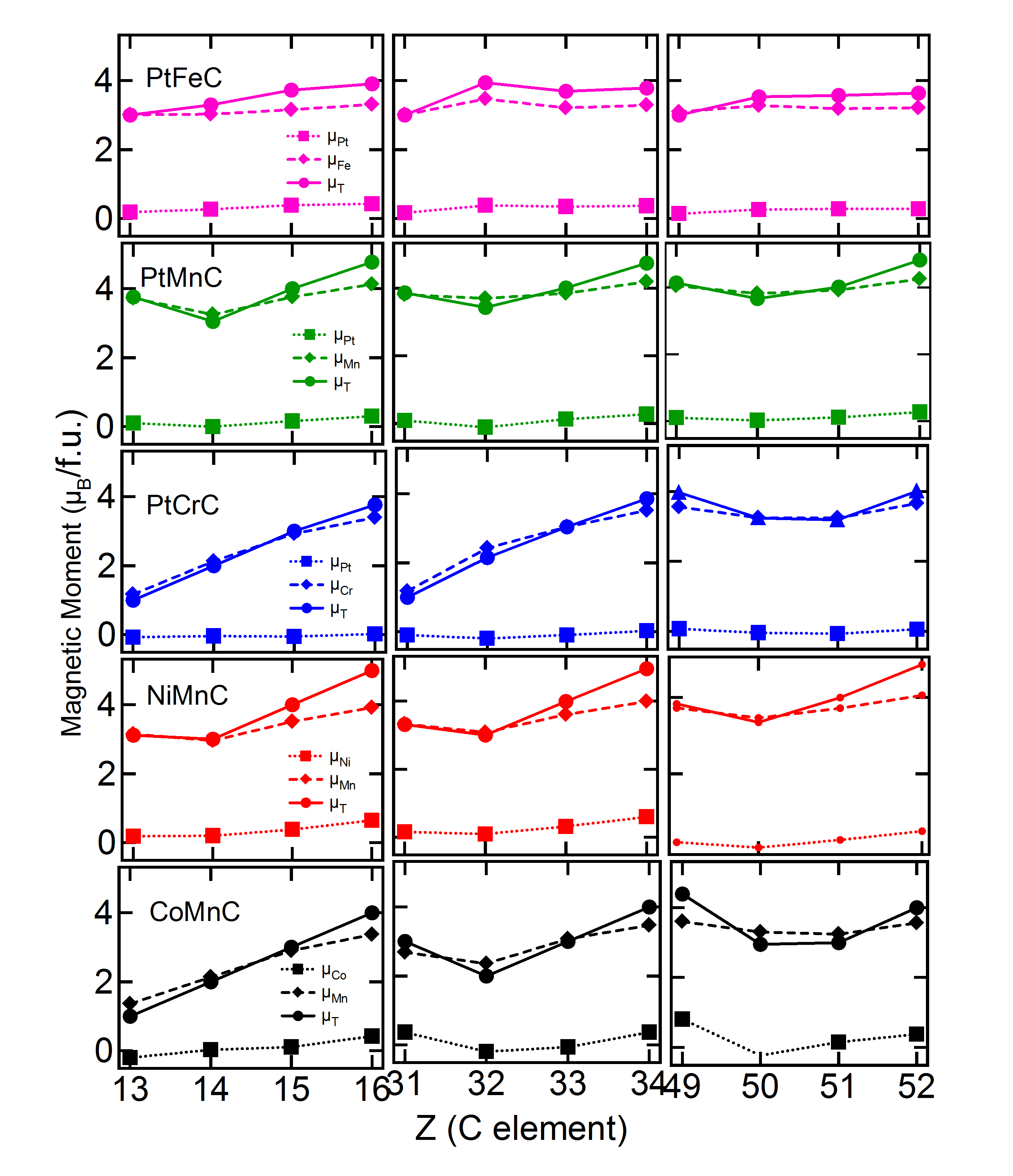}
\caption{
The total and partial moments of some of the Co$BC$, Ni$BC$ and
Pt$BC$ alloys as a function of atomic number of $C$ atoms.
}
\label{fig:2}
\end{figure}

\begin{figure}
\centering
\includegraphics[width=1.0 \textwidth]{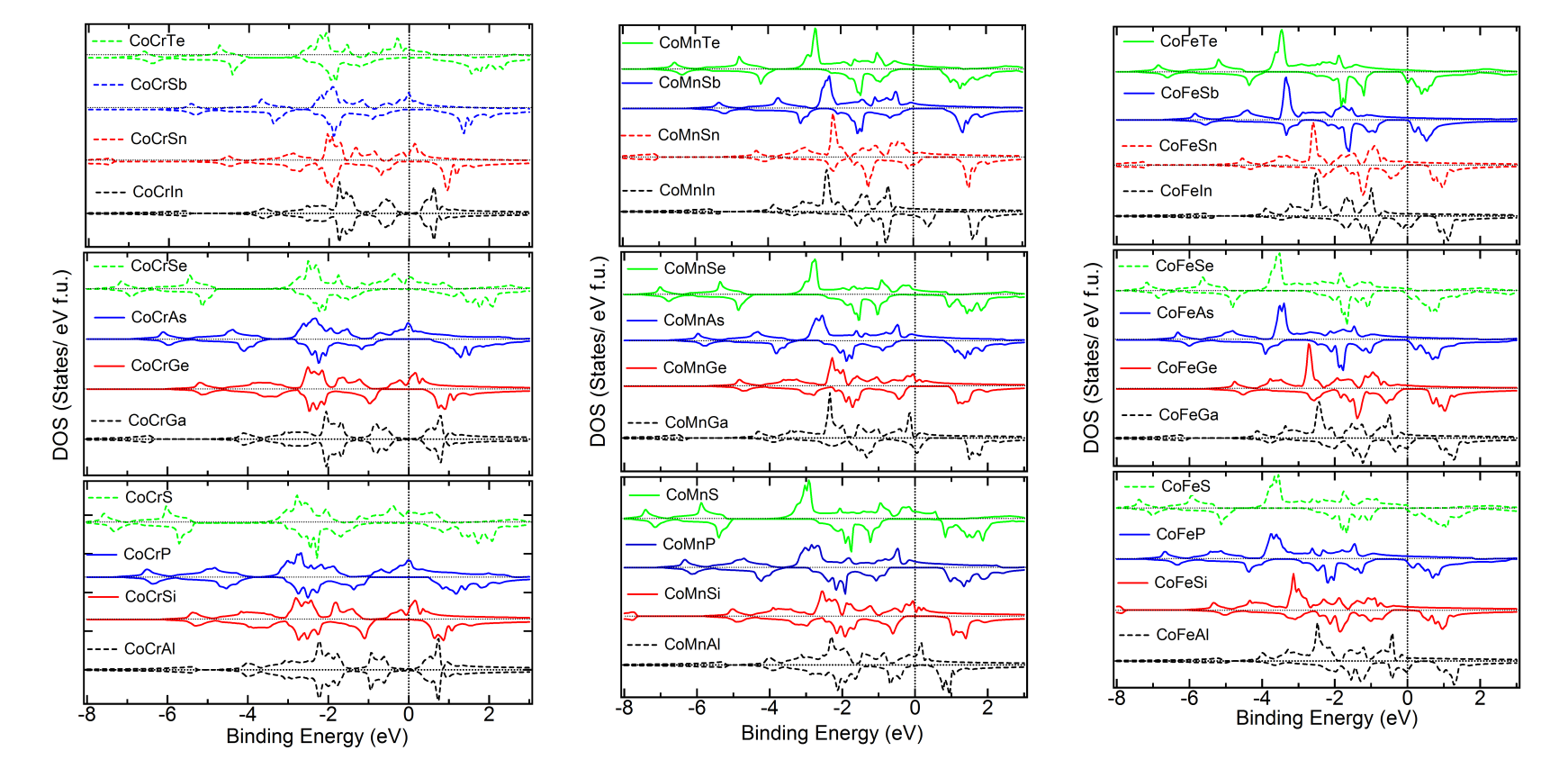}
\caption{
The up and down total density of states of Co$BC$ alloys in the
cubic phase. 
The DOS of stable alloys in terms of formation energy are  
plotted with a solid line and DOS of energetically unstable alloys
are plotted with a dotted line.
}
\label{fig:3}
\end{figure}

\begin{figure}
\centering
\includegraphics[width=1.0 \textwidth]{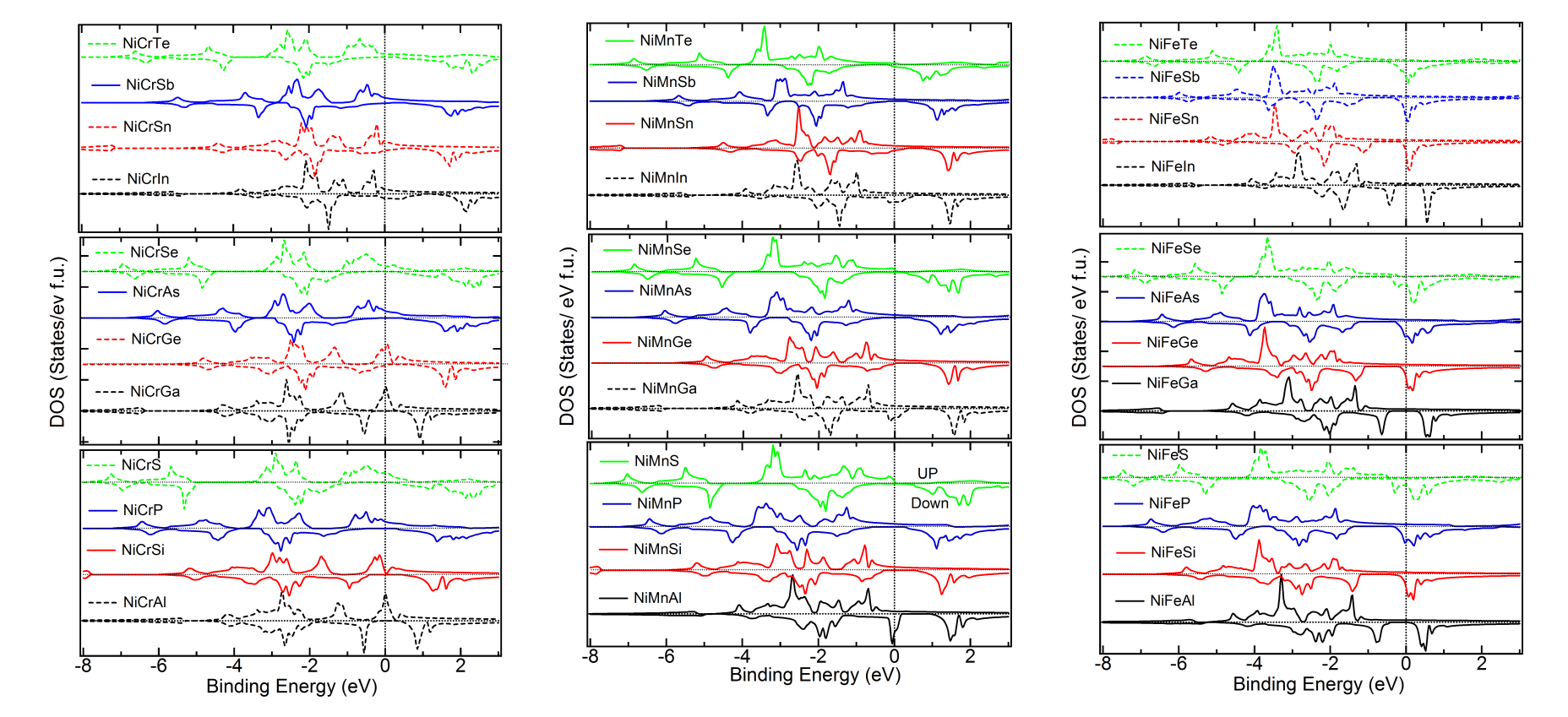}
\caption{
The up and down density of states of Ni$BC$ alloys in the
cubic phase.
The DOS of stable alloys in terms of formation energy are  
plotted with a solid (red) line and DOS of energetically unstable 
alloys are plotted with a (black) dotted line.
}
\label{fig:4}
\end{figure}

\begin{figure}
\centering
\includegraphics[width=1.0 \textwidth]{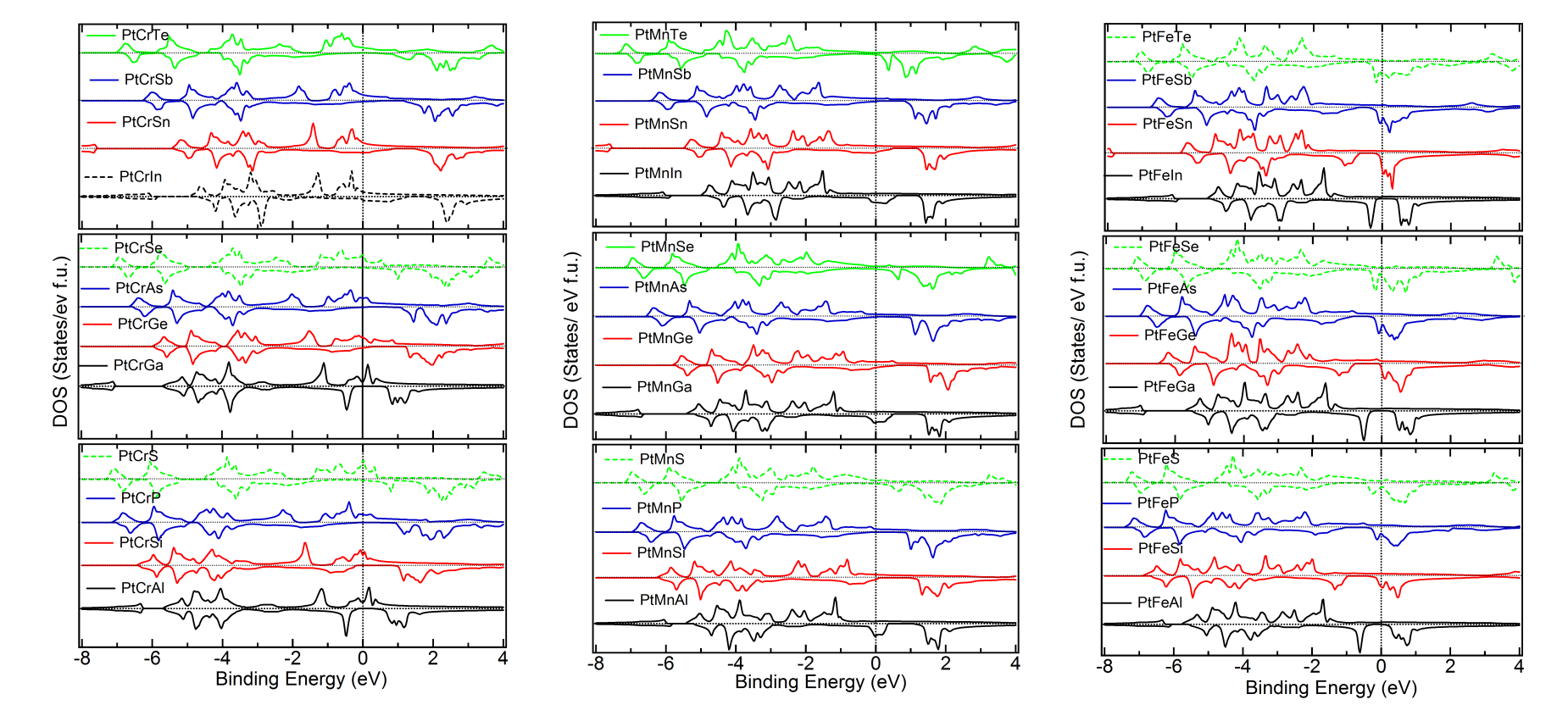}
\caption{
The up and down density of states of Pt$BC$ alloys in the
cubic phase.
The DOS of stable alloys in terms of formation energy are  
plotted with a solid (red) line and DOS of energetically unstable 
alloys  are plotted with a (black) dotted line.
}
\label{fig:5}
\end{figure}

\begin{figure}
\centering
\includegraphics[width=1.0 \textwidth]{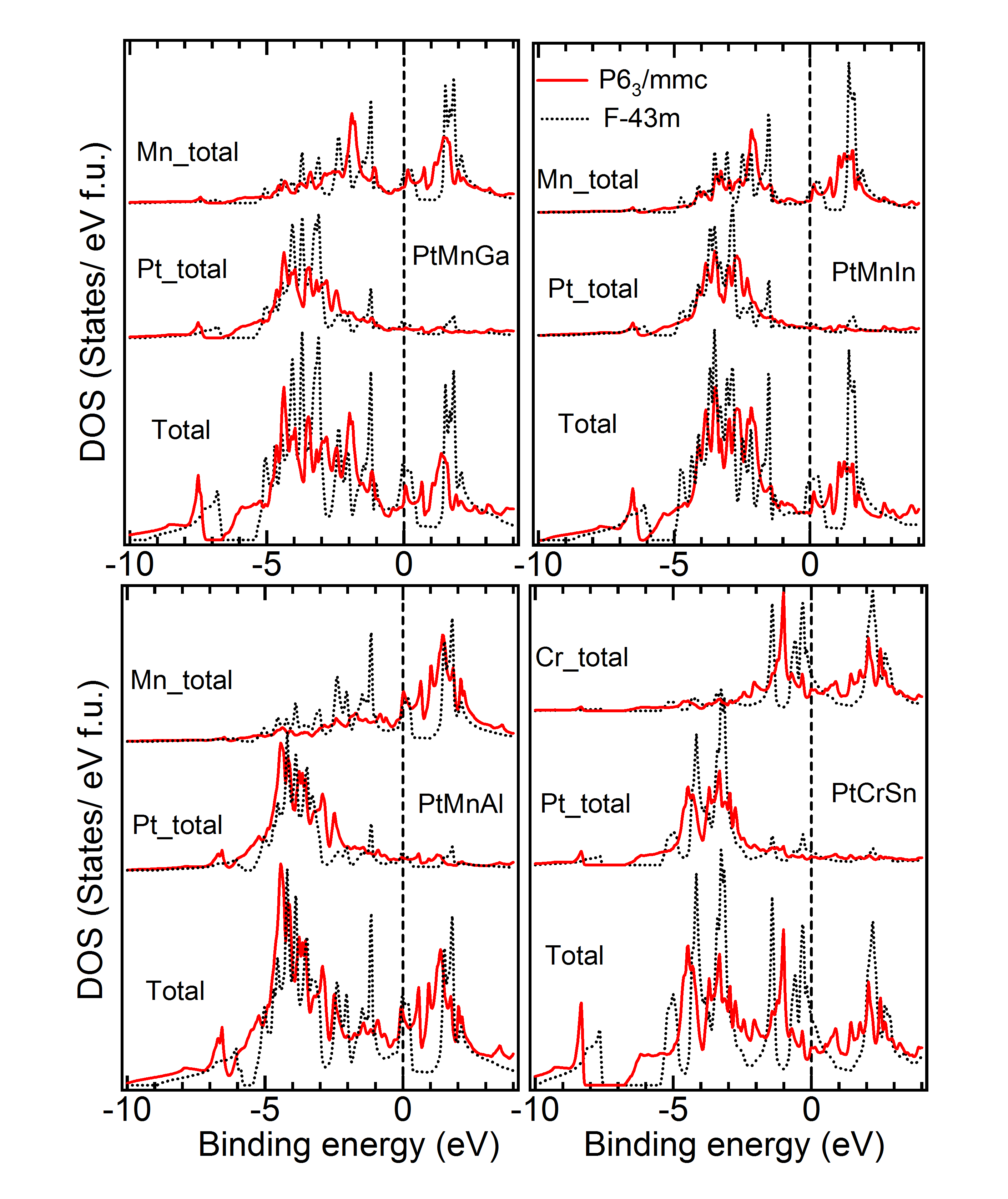}
\caption{
The total and partial DOS of the cubic and ground state
of a few materials with $P6_{3}/mmc$ symmetry.
The DOS of ground state is  
plotted with a solid (red) line and DOS of cubic  
phase is plotted with a (black) dotted line.
}
\label{fig:6}
\end{figure}
\begin{figure}
\centering
\includegraphics[width=1.0 \textwidth]{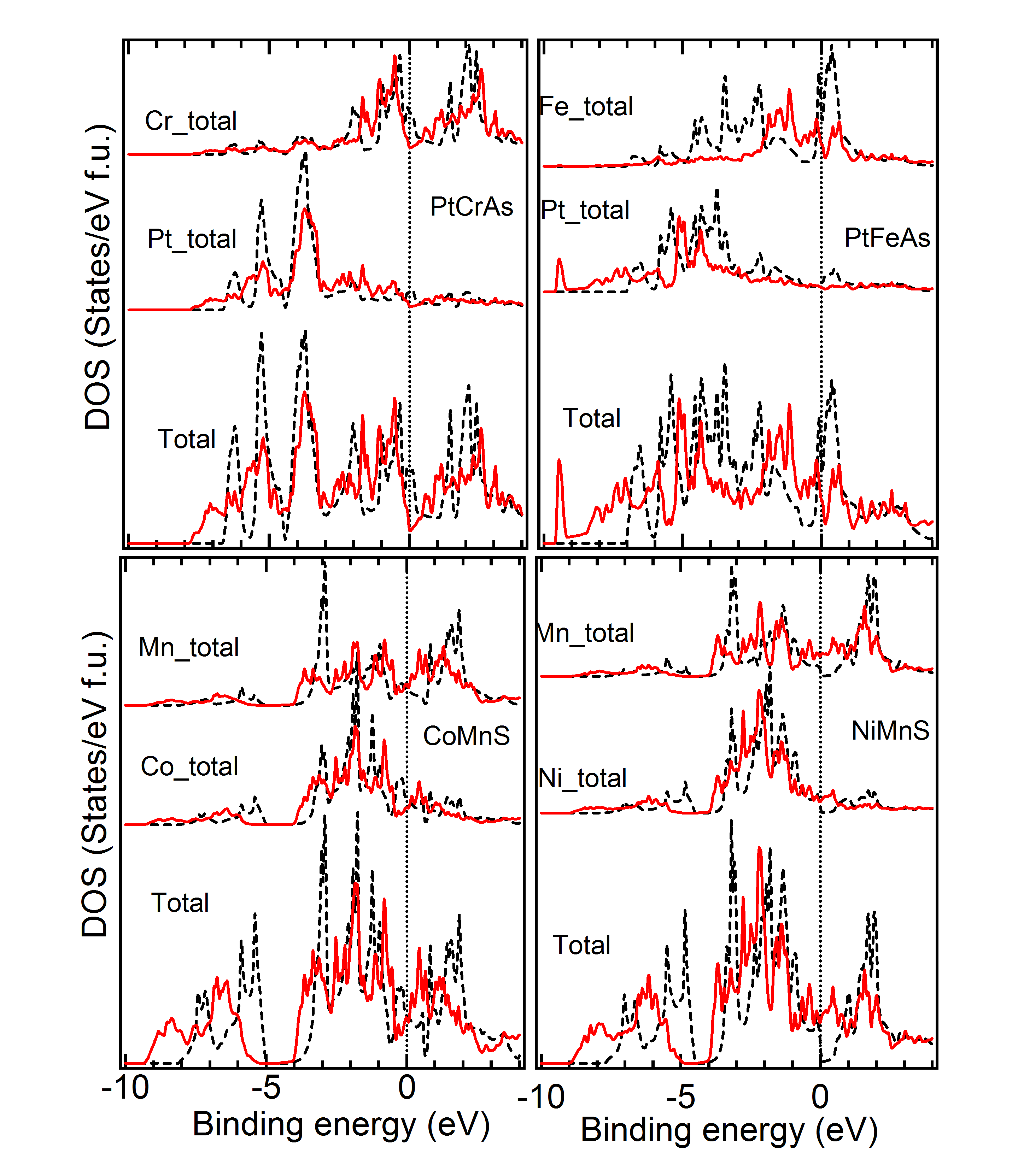}
\caption{
The total and partial DOS of the cubic and ground state
of a few materials with $P\bar{6}2m$ symmetry.
The DOS of ground state is  
plotted with a solid (red) line and DOS of cubic  
phase is plotted with a (black) dotted line.
}
\label{fig:7}
\end{figure}
\begin{figure}
\centering
\includegraphics[width=1.0 \textwidth]{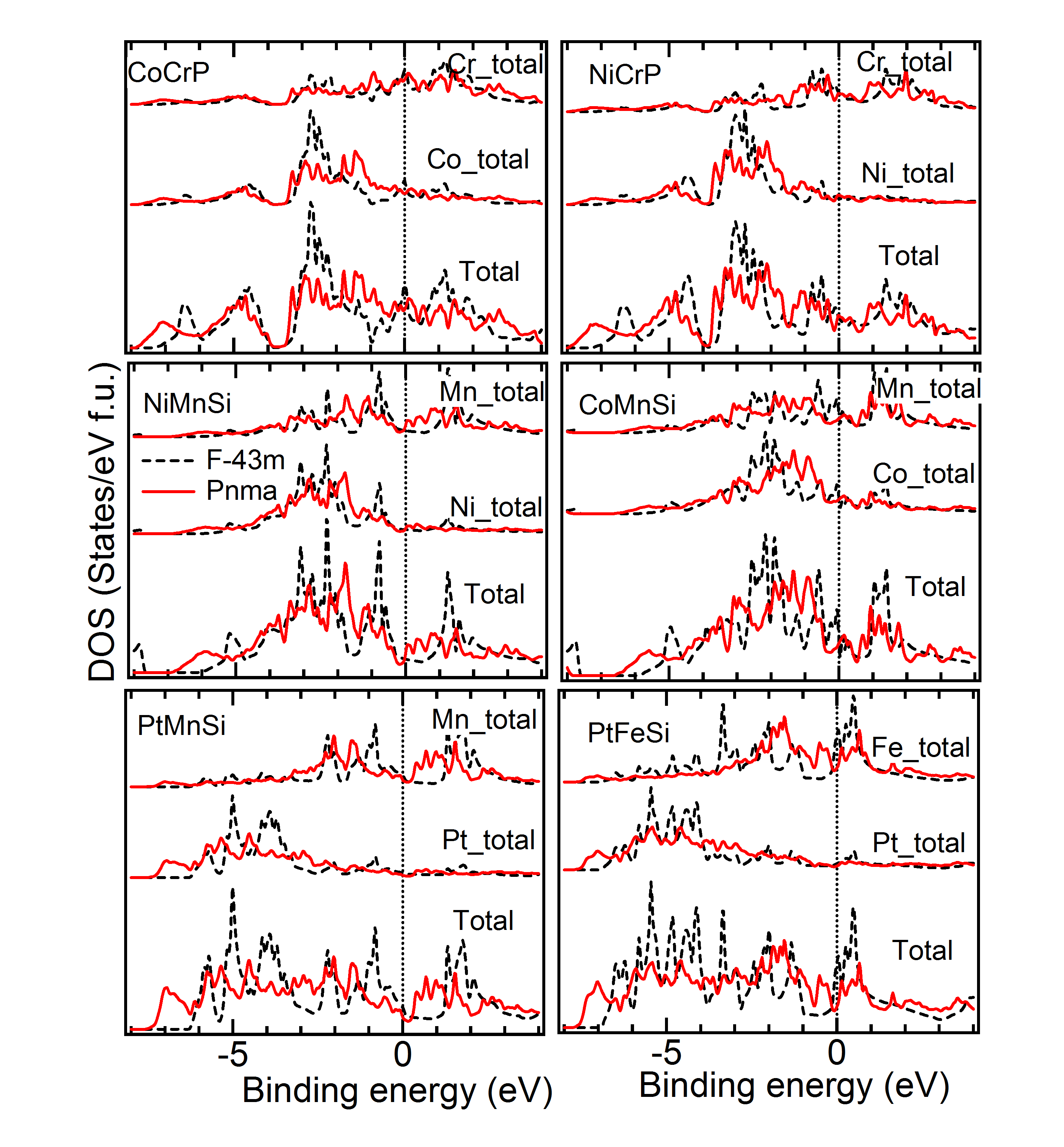}
\caption{
The total and partial DOS of the cubic and ground state
of a few materials with $Pnma$ symmetries.
The DOS of ground state is  
plotted with a solid (red) line and DOS of cubic  
phase is plotted with a (black) dotted line.
}
\label{fig:8}
\end{figure}
\begin{figure}
\centering
\includegraphics[width=1.0 \textwidth]{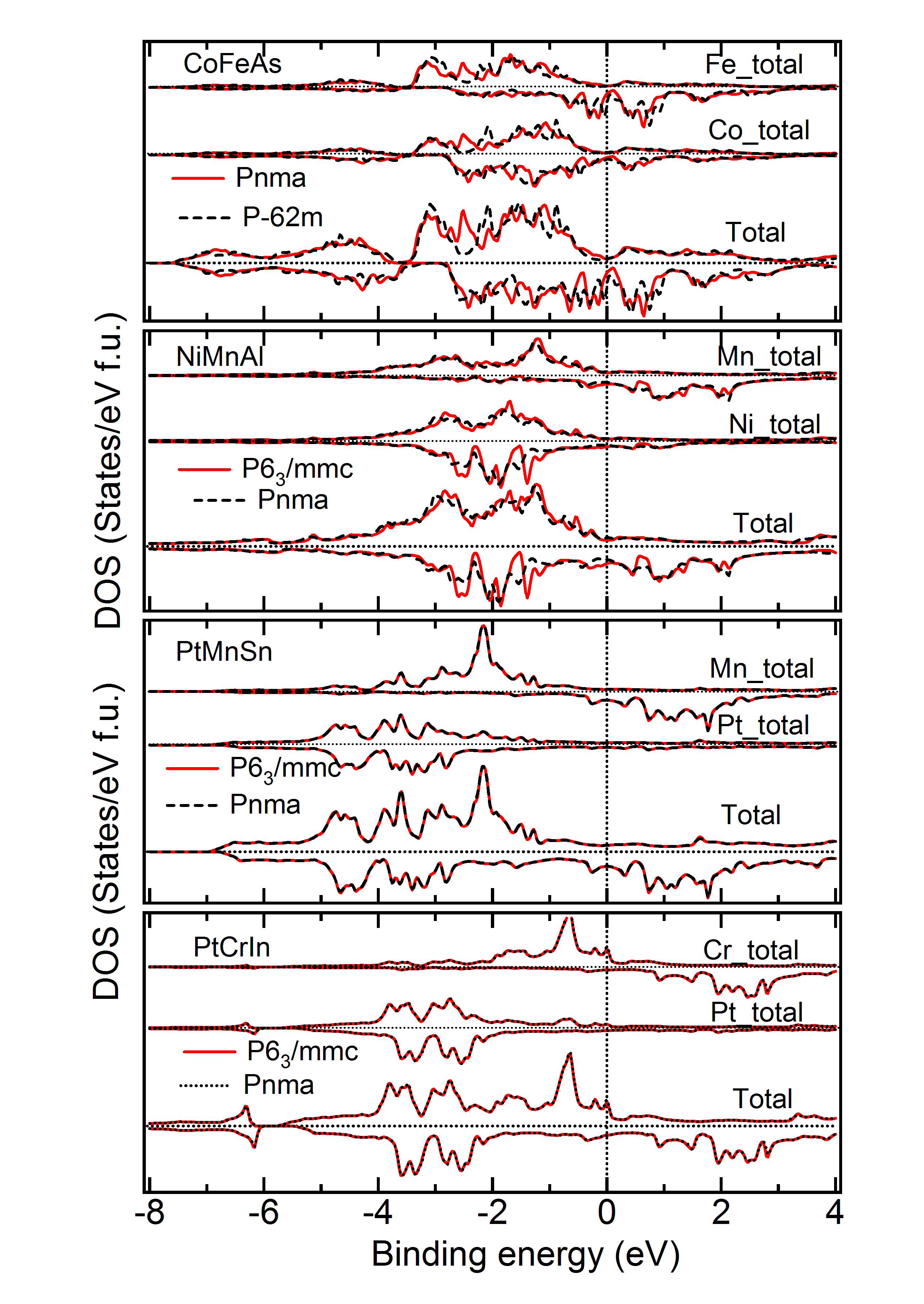}
\caption{
The total density of states and DOS of $A$ and $B$ atoms, 
for a few materials, for which two symmetries yield close values 
of $E_{form}$ and also geometry: in lowest to uppermost panels, DOS of
 both the phases for PtCrIn, PtMnSn, NiMnAl and CoFeAs are shown. 
The DOS of one phase is  
plotted with a solid (red) line and DOS of the other 
phase is plotted with a (black) dotted line.
}
\label{fig:9}
\end{figure}

\begin{figure}
\centering
\includegraphics[width=1.0 \textwidth]{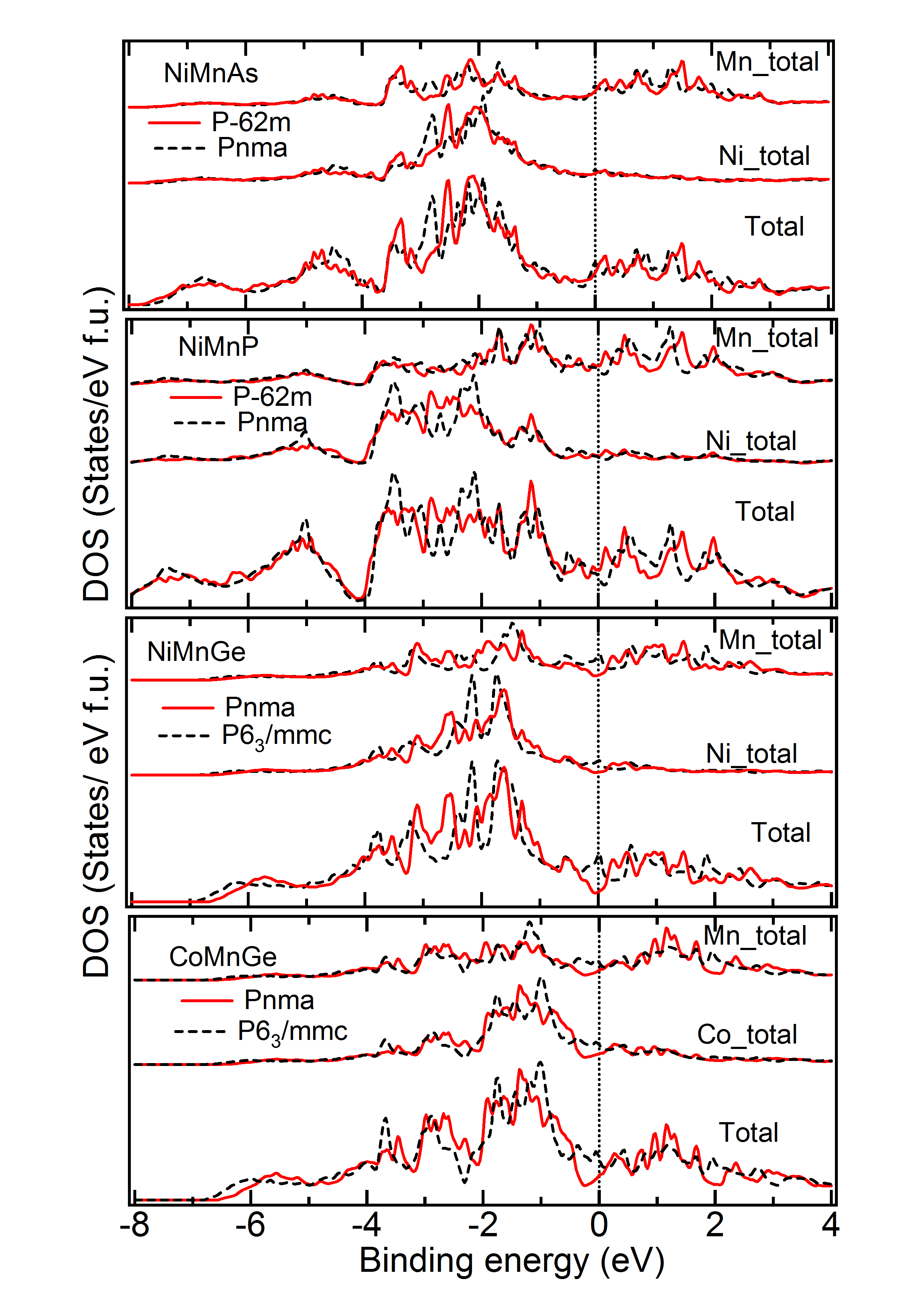}
\caption{
The total density of states and DOS of $A$ and $B$ atoms, 
for a few materials, for which experimentally two symmetries are 
reported: in lowest to uppermost panels, DOS of both the symmetries 
for CoMnGe, NiMnGe, NiMnP and NiMnAs are shown. 
The DOS of one phase is  
plotted with a solid (red) line and DOS of the other 
phase is plotted with a (black) dotted line.
}
\label{fig:10}
\end{figure}

\begin{figure}
\centering
\includegraphics[width=1.0 \textwidth]{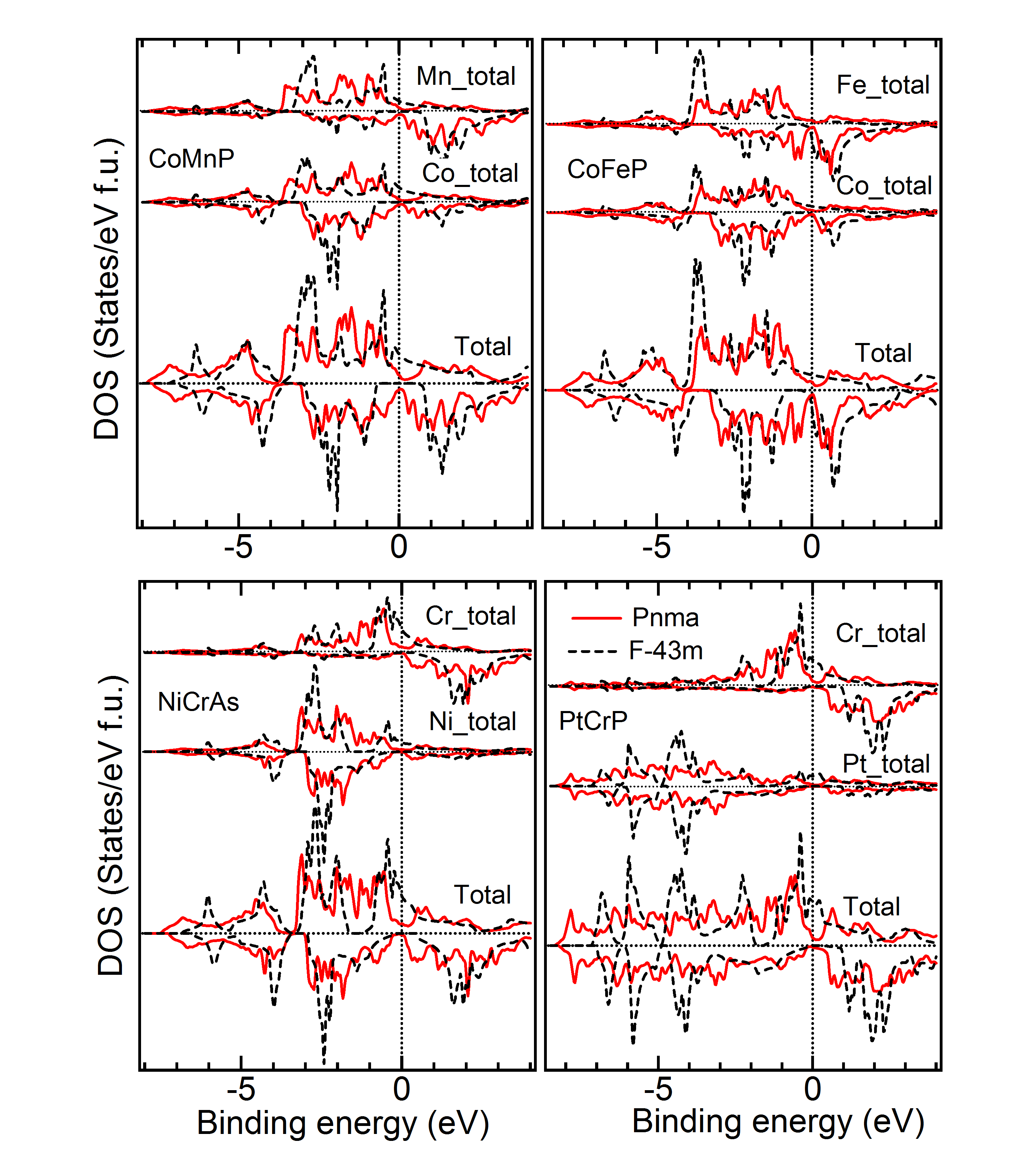}
\caption{
The total density of states and DOS of $A$ and $B$ atoms, 
for a few materials, for the non-cubic ground state and cubic phases: 
in lower panels DOS of CoMnP and CoFeP and in upper panels, 
DOS NiCrAs and PtCrP are shown. 
The DOS of one phase is  
plotted with a solid (red) line and DOS of the other 
phase is plotted with a (black) dotted line.
}
\label{fig:11}
\end{figure}

\end{document}